\documentclass[prb,preprint]{revtex4}

\pdfoutput=1

\usepackage{graphicx}

\begin{document}

\title{Coulomb-oscillator origin of superconductivity in p-doped \linebreak copper oxides

\medskip }

\date{June 27, 2013} \bigskip

\author{Manfred Bucher \\}
\affiliation{\text{\textnormal{Physics Department, California State University,}} \textnormal{Fresno,}
\textnormal{Fresno, California 93740-8031} \\}

\begin{abstract}
Emergence, development and cessation of superconductivity in three representative compounds of copper oxide families---cation doped $Ca_{2-x}Na_{x}CuO_{2}Cl_{2}$ and $La_{2-x}Ae_{x}CuO_{4}$ ($Ae = Ba, Sr$), as well as oxygen enriched $YBa_{2}Cu_{3}O_{6+x}$---are explained with the Coulomb-oscillator model of superconductivity. By the model, non-resistive current is carried by axial Coulomb oscillations of $s$ electrons through neighbor nuclei---here excited $3s$ electrons from $O^{2-}$ ions through next-nearest neighbor oxygen nuclei---if their accompanying lateral oscillation is sufficiently confined to prevent lateral overswing. Cation doping gives rise to a superlattice in the layers that sandwich each $CuO_{2}$ plane. In $Ca_{2-x}Na_{x}CuO_{2}Cl_{2}$, having one $CuO_{2}$ plane per unit cell, superconductivity emerges when laterally confined Coulomb oscillators start connecting along $6 \times 6$ superlattice domains (in units of planar lattice constants) and it peaks at $4 \times 4$ domains when, at doping $x = 1/8$, the superlattice is completed. With further doping a new, off-set superlattice grows. Its frustrating effect gradually reduces superconductivity to cessation. The same mechanism holds for $La_{2-x}Ae_{x}CuO_{4}$ which has two, staggered $CuO_{2}$ planes per unit cell. The staggering causes superconducting frustration or boost between adjacent layer sandwiches. This results in a double hump of the transition temperature $T_{c}(x)$, instead of a dome, with a deep furrow or dip at $x = 1/8$ for $Ae = Ba$ or $Sr$, respectively. Oxygen enrichment of $YBa_{2}Cu_{3}O_{6+x}$ indirectly leads to effective doping in the $CuO_{2}$ planes themselves ($Cu^{2+} \to Cu^{3+}$). The ionization of copper ions at the corners of planar unit cells determines whether lateral oscillations between next-nearest neighbor $O^{2-}$ ions overswing ($T_{c} = 0$) or are confined to wide or narrow electron tracks. Their percolating connectivity gives rise to respective plateaus of $T_{c} \simeq 57$ K and $T_{c} \simeq 90$ K, and intermediate ramps.

\end{abstract}

\maketitle


\section{INTRODUCTION}
Superconductivity emerges, develops and ceases in a class of copper oxides depending on cation doping or oxygen enrichment. Here that dependence is explained with the Coulomb-oscillator model of superconductivity.\cite{1}  Representative compounds of three copper-oxide families will be discussed where the $CuO_{2}$ planes occur (1) as a single plane in the unit cell, $Ca_{2-x}Na_{x}CuO_{2}Cl_{2}$, (2) as staggered double planes, $La_{2-x}Ae_{x}CuO_{4}$ with alkaline-earth dopants $Ae = Ba$ or $Sr$, and (3) as aligned double planes, $YBa_{2}Cu_{3}O_{6+x}$. After some general considerations which concern all three cuprate families, the single-plane case provides the best opportunity to explore a superlattice imposed by $Na$ doping of the sandwiching $CaO$ layers.  The superlattice concept carries over to the compound family with staggered double planes, also doped by cation substitution. The stark contrast with the single-plane case results from the staggering, while an internal distinction between the sister compounds originates with structural details and the different size of doping ions, $Ae^{2+} = Ba^{2+}$ or $Sr^{2+}$. In the case of $YBa_{2}Cu_{3}O_{6+x}$, oxygen enrichment indirectly gives rise to an effective doping in the $CuO_{2}$ planes themselves ($Cu^{2+} \to Cu^{3+}$). Here superconductivity is dominated by the ionization of the planar unit areas as they aggregate to domains.

\section{GENERAL CONSIDERATIONS}
The semiclassical Coulomb-oscillator model of superconductivity attributes non-resistive electron current exclusively to the motion of the atoms' outer $s$ electrons through the nuclei of neighbor atoms (axial Coulomb oscillation).  Superconductivity is destroyed if the electrons' accompanying lateral oscillation, halfway at and perpendicular to the internuclear axis, extends between neighbor atoms. (The model's neglect of the electrons' wave character prevents the notion of connecting both oscillation modes in a continuous process. However, one can imagine that on average each electron spends half the time in either axial or lateral oscillation.) The third tenet of the Coulomb-oscillator model of superconductivity is the \emph{squeeze effect}, whereby a compression of the atoms that provide the relevant outer $s$ electrons causes a reduction of lateral oscillation width.  The difference $\Delta y$ between the lateral oscillation width, $B0B$, and the lateral confinement limit, $\overline{B0B}$, is a proportional measure for the transition temperature of superconductivity, $\overline{B0B} - B0B \equiv \Delta y \propto T_{c}$.\cite{1}

\includegraphics[width=6in]{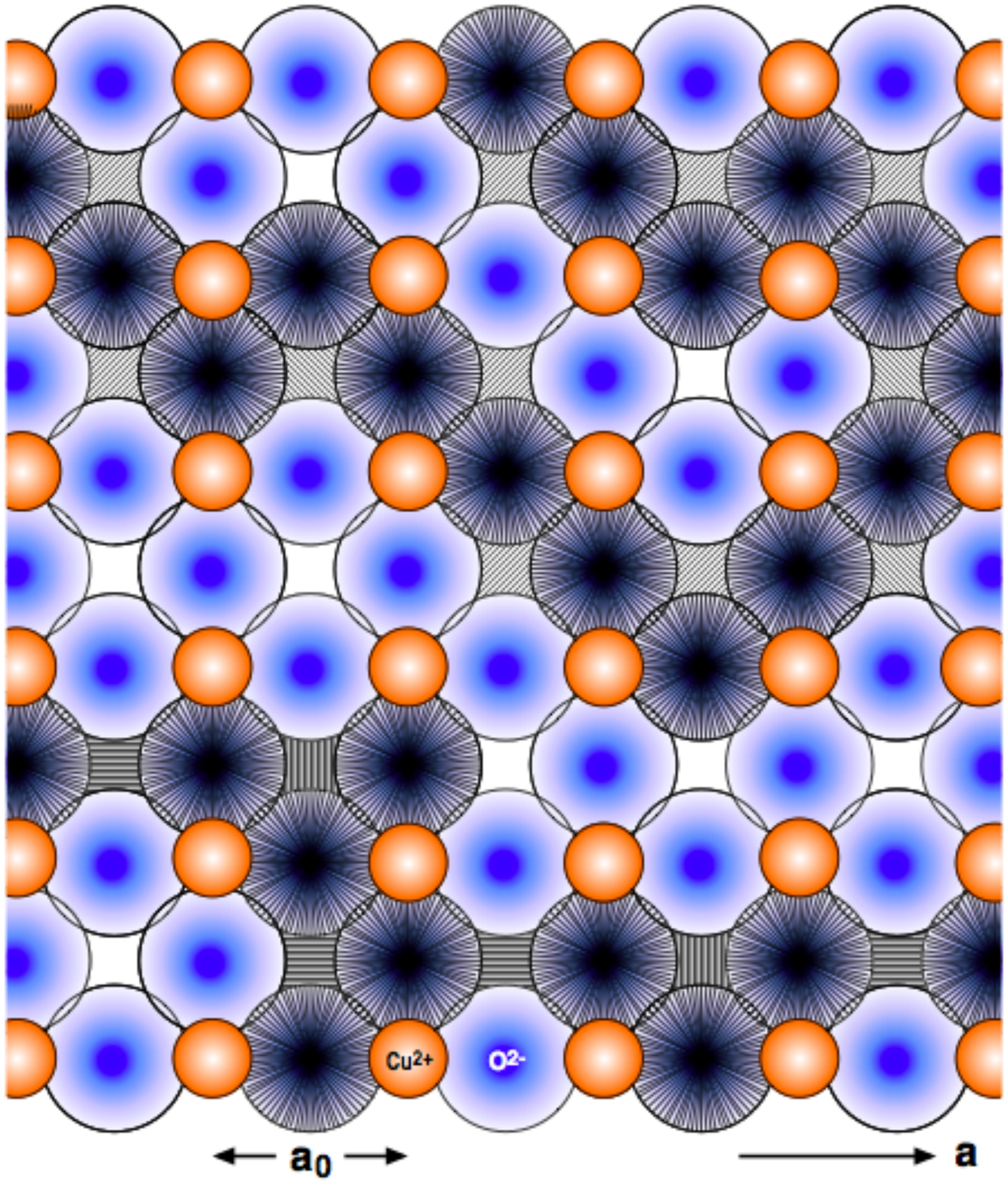}

\noindent FIG. 1. Schematic time sequence of two diffusing $3s$ electrons in the $CuO_{2}$ plane of an undoped cuprate compound due to  axial and lateral overswing between nearest-neighbor oxygen sites (top) and between next-nearest-neighbor oxygen sites (bottom).
\pagebreak

It is widely believed that high-temperature superconductivity of the copper oxides occurs in the $CuO_{2}$ plane. Neither a free $Cu^{2+}$ ion nor an $O^{2-}$ ion in an alkaline-earth oxide---free $O^{2-}$ ions are unstable---possesses an outer $s$ electron as required by the Coulomb-oscillator model. However, in the compounds under consideration the strong interionic Coulomb forces exert a compression of each $CuO_{2}$ layer as a whole as well as of the $O^{2-}$ ions within that layer by their neighbor $Cu^{2+}$ ions (and also by $Cu^{3+}$ in the case of $YBa_{2}Cu_{3}O_{6+x}$). In reaction, the crystal reduces short-range repulsion between the $Cu^{2+}$ and $O^{2-}$ ions by an electron reconfiguration in $O^{2-}$ from $2p^{6}$ to $2p_a^1p_b^1p_c^23{s^2}$, that is, by reducing the occupancy of the $2p_{a}$ and $2p_{b}$ orbitals that are aligned along the $O$-$Cu$ axes in favor of excited, isotropic $3s$ orbitals.  The oxygen ions then have uncompensated spins and $\ell  = 1$ orbital angular momenta, with an associated combined magnetic moment $\textbf{m}_{O}$. Both the excited $3s$ electrons and the magnetic moments $\textbf{m}_{O}$ have profound consequences on properties of the copper-oxide compounds. Only the electronic aspect, $3s$, will be discussed here as it affects superconductivity.

The excited $3s$ electrons of extremely squeezed $O^{2-}$ ions participate in Coulomb oscillations both between nearest-neighbor ($nn$) $O^{2-}$ ions (diagonally to the crystal's $a$ and $b$ axes) and between next-nearest-neighbor ($nnn$) $O^{2-}$ ions---those without intervening $Cu^{2+}$ ions---parallel to the $a$ or $b$ axis. Because of their lesser nuclear separation, and being twice as numerous, the Coulomb oscillations of the $nn$ case dominate over those of the $nnn$ case.  Another distinction is that the \emph{lateral} oscillations that accompany the Coulomb oscillations of $3s$ electrons between $nn$ $O^{2-}$ ions \emph{always} overswing, not only in the undoped crystal but also in the doped compounds, treated below. Those $3s$ electrons diffuse through the $CuO_{2}$ plane by random axial or lateral overswing. Such diffusion scenarios are schematically depicted in Fig. 1 for two $3s$ electrons---for sake of demonstration, one only by Coulomb oscillations between $nn$ oxygen pairs, the other between $nnn$ pairs. In reality combinations of both cases occur.

In the framework of band theory the delocalized $3s$ electrons in the $CuO_{2}$ plane fill quantum states in the reciprocal lattice ($k$ space). In terms of energy, the filled $3s$ band resides below the chemical potential $\mu$ and is separated by an energy gap from a higher empty band.  That's why undoped $Ca_{2}CuO_{2}Cl_{2}$ and $La_{2}CuO_{4}$, as well as $YBa_{2}Cu_{3}O_{6}$, are insulating instead of metallic.

\pagebreak
\includegraphics[width=5.95in]{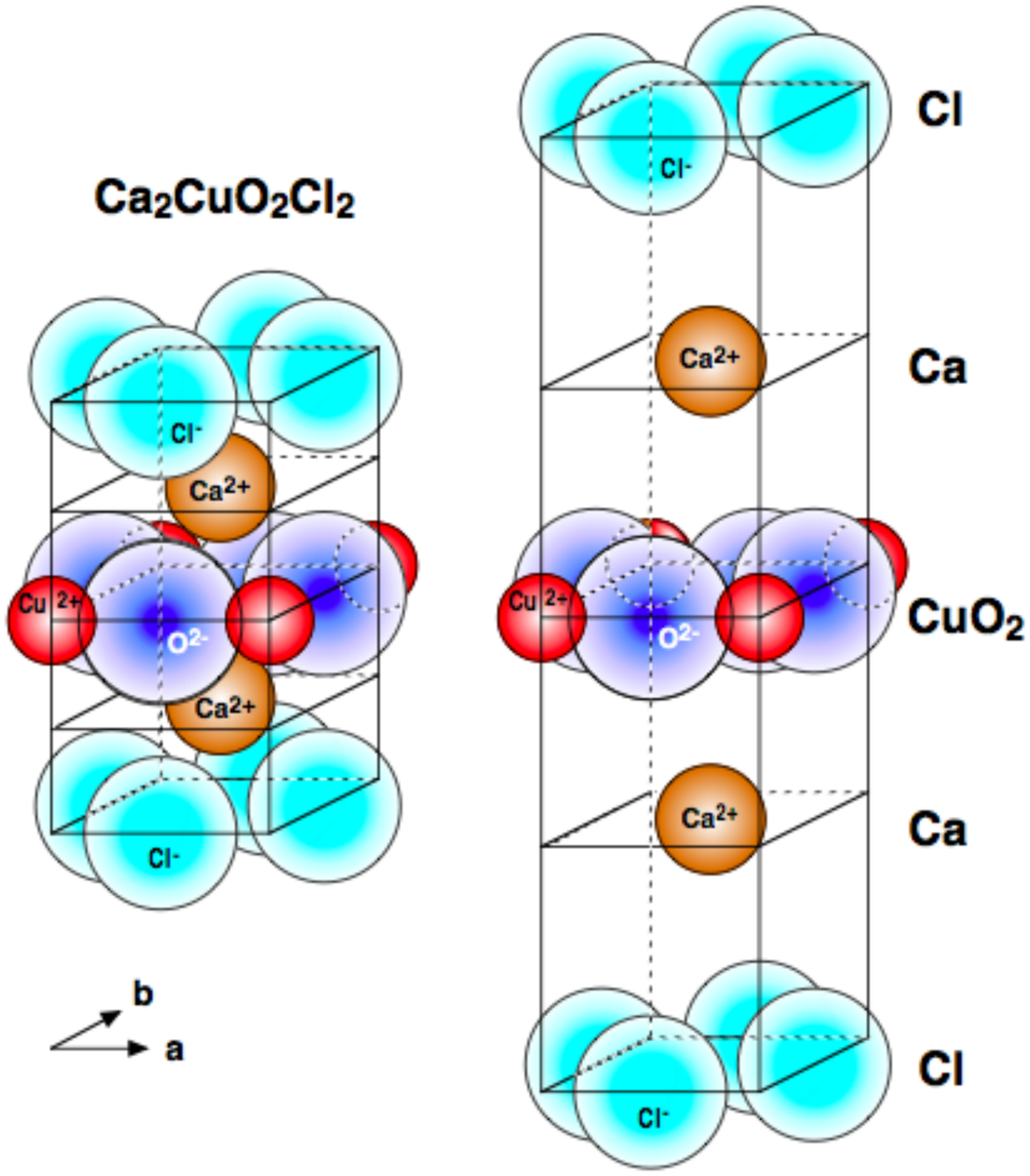}

\noindent FIG. 2.  Unit cell of $Ca_{2}CuO_{2}Cl_{2}$ in hard-sphere display (left) and vertically exploded (right).
\pagebreak

\section{SUPERCONDUCTIVITY IN $\mathbf{Ca_{2-x}Na_{x}CuO_{2}Cl_{2}}$}

Figure 2 shows the crystal structure of undoped $CaCuO_{2}Cl_{2}$.  Its unit cell contains a single $CuO_{2}$ plane, sandwiched above and beneath by a plane of $Ca^{2+}$ cations, which in turn are covered above or beneath by a plane of $Cl^{-}$ anions. Sodium doping substitutes $Na^{+}$ ions for $Ca^{2+}$ in the sandwiching cation layers, denoted as $Ca_{2-x}Na_{x}CuO_{2}Cl_{2}$. It is observed that the antiferromagnetic phase of the undoped crystal is destroyed for doping $x > 0.02$. In the \emph{underdoped} range, $0 < x < 0.125 = 1/8$, a square superlattice of spacing
\begin{equation}
{A^ + }(x) = \sqrt {\frac{2}{x}} \; {a_0}
\end{equation}

\noindent evolves in the $CuO_{2}$ plane, determined by vertical $Na^{+}$ \emph{pairs} (parallel to the crystal's $c$ axis) residing in the cation layers above and beneath. 

For a familiarization with domain geometry, consider the frame in Fig. 3. It contains a $4a_{0} \times 4a_{0}$ domain when the doping is $x = x_{4} \equiv 1/8$.  The positions of the vertical $Na^{+}$ pairs are marked each by a cross ($\textbf{+}$) at the middle of the left and right frame side. This particular arrangement of the fame, relative to the $Na^{+}$ pairs, will be advantageous for an explanation of the doping dependence of superconductivity, $T_{c}(x)$. But in order to see clearly the relation of the $Na^{+}$-pair superlattice and square domains, imagine briefly the frame shifted downward (parallel to the negative $b$ axis) by $2a_{0}$.  The $Na^{+}$ pairs then fall on the corners of the shifted frame and demark a planar unit cell of the superlattice, here ${{A^ + }^2}(x_{4}) = {(4{a_0})^2}$.

For the relevant range of underdoping, $0.02 < x < 0.125$, this corresponds to domain-size shrinking from $10{a_0}$ to $4{a_0}$. In fact, the destruction of the antiferromagnetic ground state at $x = x_{10} \equiv  0.02$ occurs when the superlattice spacing is $A^{+} = 10a_{0}$. Preservation of crystal symmetry requires a discrete spacing of $Na^{+}$ pairs at (or very close to) integer multiples of the lattice constant, $A_{n}^{+} = n a_{0}$, that is, $n = 10,9,...,4$, for the non-antiferromagnetic \emph{underdoped} range. As we'll see, many compound properties of  $Ca_{2-x}Na_{x}CuO_{2}Cl_{2}$ show a continuous doping dependence.  How can that be understood in terms of discrete domain sizes, $A_n^+$? In this context the continuous domain size $A^{+}(x)$, Eq. (1), should be regarded as the spatial average of discrete domains over a large area.  For example, $A^{+}(x=5.5)$ would correspond to an area of alternating $A_6^+$ and $A_5^+$ domains.

The different charge of the dopant ions $Na^{+}$ from the replaced host cations $Ca^{2+}$ gives 

\pagebreak
\includegraphics[width=5.05 in]{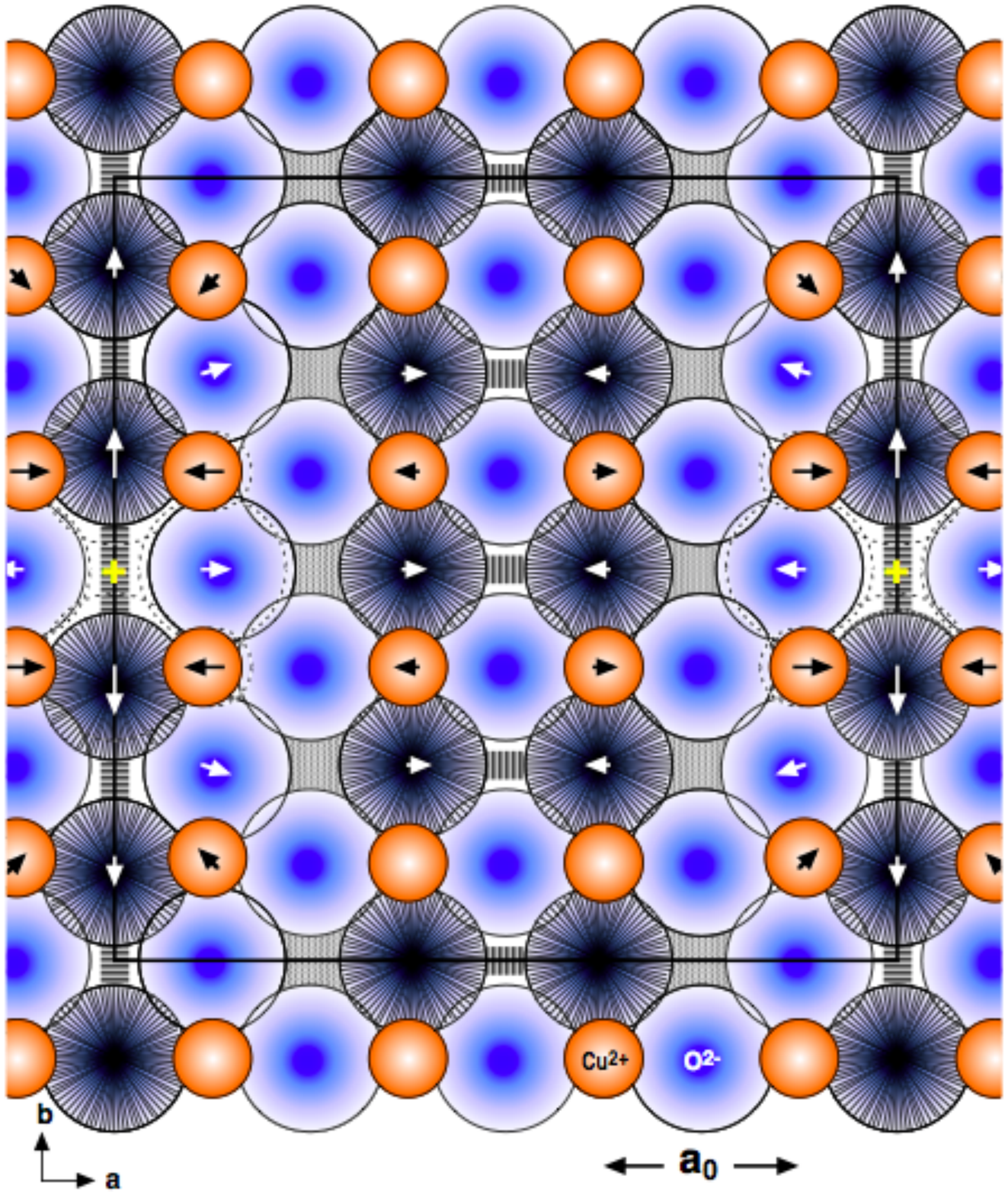}

\noindent FIG. 3. $CuO_{2}$ plane of $Ca_{2-x}Na_{x}CuO_{2}Cl_{2}$, $x = 0.125$, with $4{a_0} \times 4{a_0}$ domain (delimited by the frame).  The crosses halfway up the sides of the frame each mark the position of a doped $Na^{+}$ ion pair residing in the plane above and beneath.  Ions are shown by their hard-sphere size.  Arrows indicate prominent ion displacements.  Dashed circles show positions in the undoped crystal for some ions. Straight-line hatching indicates Coulomb oscillation of $3s$ electrons through oxygen nuclei and, laterally confined, across the interionic region. Shaded interionic space signifies lateral $3s$ overswing. The combined complex of an $Na^{+}$ pair and unidirectional Coulomb oscillators along the right or left frame side comprises an \emph{``epaulet,''} $E^{+}$, here overlapping with corresponding epaulets of the upper and lower neighbor domains. 

\pagebreak

\noindent rise to displacement of nearby ions in the $CuO_{2}$ plane. A trend well known from lattice defects in ionic crystals, close $O^{2-}$ ions are displaced \emph{away} from each vertical pair of substitutional $Na^{+}$ ions ($\textbf{+}$ mark on the domain frame in Fig. 3).  Somewhat counterintuitive with regard to space availability in terms of ionic radii---$r(Ca^{2+}) = 0.99$ {\AA}, $r(Na^{+}) = 0.95$ {\AA}---the reason is that the accumulate long-range Coulomb forces dominate over interionic short-range repulsion.  For a simplified explanation, one can think that the outward displacement of neighboring $O^{2-}$ ions results from their lesser electrostatic attraction to $Na^{+}$ than to $Ca^{2+}$.  Likewise the inward displacement of nearby $Cu^{2+}$ ions is caused by their lesser electrostatic repulsion from $Na^{+}$ than from $Ca^{2+}$.  

There are two equilibrium configurations of displaced $O^{2-}$ ions:  The configuration with fourfold symmetry, dominated by \emph{equal} outward displacements of the four $O^{2-}$ ions that are nearest neighbors to an $Na^{+}$ pair, is unstable (not shown in Fig. 3).  Any slight disturbance would cause a transition to the stable equilibrium configuration of \emph{twofold} (mirror) symmetry and presumably lower energy. This is the root of the much-quoted symmetry breaking in the superconducting copper oxides. It is dominated by \emph{unequal} displacements of the four closest $O^{2-}$ ions away from the $Na^{+}$ pair---large outward displacements of two opposite closest $O^{2-}$ neighbors (white arrows up and down in Fig. 3) and small outward displacements of the other two closest $O^{2-}$ neighbors (to the left and right of the fame sides in Fig. 3). The displacement-induced, \emph{extra} squeeze of some $O^{2-}$ ions has profound consequences on superconductivity.

The lateral oscillations that accompany the Coulomb oscillations of excited $3s$ electrons
between $nnn$ $O^{2-}$ ions overswing in the undoped compounds, as they always do between $nn$ $O^{2-}$ ions. However, the \emph{extra} squeeze of some $O^{2-}$ ions, caused by the large displacements of $Cu^{2+}$ ions near dopant $Na^{+}$ pairs, sufficiently reduce the lateral width of those $3s$ Coulomb oscillation between the corresponding $nnn$ oxygen nuclei.  Accordingly such $3s$ electron oscillations are \emph{localized} to unidirectional electron tracks, depicted in Fig. 3, instead of diffusing throughout the $CuO_{2}$ plane.

Most important are the laterally confined Coulomb oscillators along two opposite sides of the domain frame (right and left sides in Fig. 3). We want to call, for short, the combined complex of an $Na^{+}$ pair and the confined unidirectional Coulomb oscillators at one frame side an \emph{``epaulet,''} denoted by $E^{+}$. The decrease of ion displacement, and accordingly of extra $O^{2-}$ squeeze, farther away from the $Na^{+}$ pair gives rise to an effective epaulet length $L_{E} = 6a_{0}$. Therefore, at the doping ratio $x_{6} \simeq 0.056$, that is, when the superlattice spacing is ${A^+} = 6a_{0}$, such Coulomb-oscillator epaulets along the left and right sides of $6{a_0} \times 6{a_0}$ domains start connecting, parallel to the $b$ axis, with epaulets of adjoining domains. This establishes percolating pathways through the $CuO_{2}$ plane, causing the onset of superconductivity.  With increased doping $x$ the domain size shrinks according to Eq. (1), providing more robust epaulet connections. The \emph{most} robust epaulet connection is reached at $x = x_{4} \equiv 1/8$. The corresponding epaulet pattern can be seen at, say, the right frame side in Fig. 3. It is the result  of the $6a_{0}$-long epaulet about the right $Na^{+}$ pair (extending by $1a_{0}$ into each the upper and lower $4a_{0} \times 4a_{0}$ neighbor domain in the $ \pm b$ direction) and the $1a_{0}$-overlap with the epaulets from those neighbor domains into the frame. The Coulomb-oscillator epaulets and their connections are destroyed by thermal agitation (lattice vibration) of energy $\varepsilon $ $ = \hbar \omega  > k_{B}{T_c}$, were $ \omega$ denotes the relevant (angular) phonon frequency, $k_{B}$ Boltzmann's constant, and $T_{c} = T_{c}(x)$ the doping-dependent transition temperature of superconductivity.  

Since the potential well in the epaulet region is more attractive than in an undoped crystal, $3s$ electrons that laterally overswing owing to thermal assistance, still hover about the epaulet region rather than diffuse throughout the $CuO_{2}$ plane (that is, unlike the illustration in Fig. 1). Lost to superconductivity, they still maintain much of the epaulets' unidirectional symmetry which can be observed at $T > T_{c}$ in terms of charge stripes.  In this view some of the charge-stripe order can be regarded as ``spilled $3s$ epaulets.''

A simple model quantifies the concept of epaulet connectivity by epaulet overlap in
 adjoining domains, $E^{+}$-$E^{+}$, defined as
\begin{equation}
\vartheta (x) \equiv \frac{{L_{E} - {A^ + }(x)}}{{2{a_0}}}, \; {x_6} \, \le \, x \, \le \, {x_4}.
\end{equation}
\noindent Normalized to unity for maximum overlap at $x = x_{4}$, epaulet connectivity emerges when the domain size attains the epaulet length, $A^{+}(x_{6}) = L_{E} = 6a_{0}$, and $\vartheta (x)$ increases with
shrinking with domain size $A^{+}(x)$ for increased doping. The graph of $\vartheta (x)$ in Fig. 4a shows the ascending arc of a cusped dome for the underdoped regime, $0 < x < x_{4}$. 

We obtain a relation of epaulet connectivity with the transition temperature of superconductivity by the assumption of a compound-dependent robustness factor $\rho$,
\begin{equation}
T_{c}(x) = \rho \; \vartheta (x).
\end{equation}

The connected $E^{+}$-$E^{+}$ epaulets, looking like sidepieces of a ladder, can be made visible 

\pagebreak
\includegraphics[width=6in]{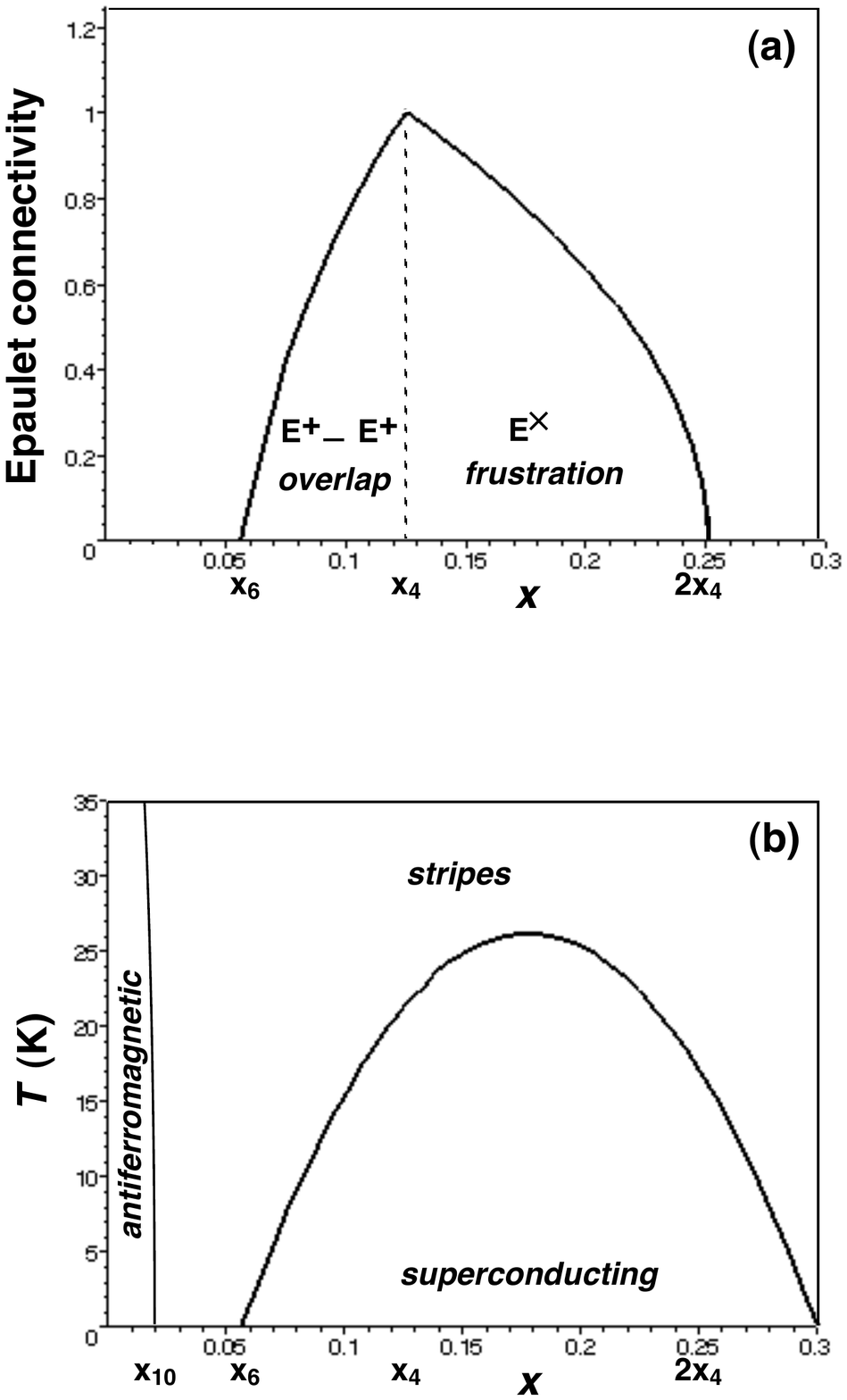}

\noindent FIG. 4. (a) Doping dependence of epaulet connectiviy $\vartheta (x)$ by $E^{+}$-$E^{+}$ overlap in the underdoped regime and by $E^{\times}$ frustration in the overdoped range. (b) Phase diagram of $Ca_{2-x}Na_{x}CuO_{2}Cl_{2}$ as experimentally observed.
\pagebreak

\noindent with asymmetric scanning tunneling microscopy (STM).\cite{2,3} They are found to be in qualitative agreement with the pattern in Fig. 3. Also visible in the STM images and in Fig. 3 are isolated local Coulomb oscillators farther away from the dopant $Na^{+}$ pairs---the rungs of the ladder---that don't contribute to superconductivity.\cite{4}

The ${A^ + }(\frac{1}{8}) \times {A^ + }(\frac{1}{8})$ superlattice, being completed, persists in the subsequent \emph{overdoped} range, $1/8 < x \le 1/4$. Moreover, a new, additional superlattice ${A^ \times }(x) \times {A^ \times }(x)$ evolves, cornered by the positions of new $Na^{+}$ pairs, with superlattice constant 
\begin{equation}
{A^ \times }(x) = \sqrt {\frac{2}{{x - _8^1}}} \; {a_0} \; .
\end{equation}
\noindent Its coordinate origin is displaced from that of the persisting superlattice, $A^{+}(\frac{1}{8}) \times A^{+}(\frac{1}{8})$, by $2a_{0}$ in both the $a$ and $b$ directions. The destructive effect of the new superlattice on superconductivity can best be seen at the far end of the doping regime, $x=0.25$, illustrated in Fig. 5. In that case the ${A^ \times }$ superlattice has shrunk to the same size as the persisting $A^{+}$ superlattice, ${A^ \times }(\frac{1}{4}) = {A^ + }(\frac{1}{8} \, \le \, x \, \le \, \frac{1}{4}) \equiv 4{a_0}$. Doped $Na^{+}$ pairs reside now not only at the sides of the domain frame (upright crosses, $\textbf{+}$, in Fig. 5) but also at the top and bottom of the frame (slanted crosses $\mathbf{\times}$).   

Compensation of forces from $Na^{+}$ pairs at both $\textbf{+}$ and $\mathbf{\times}$ positions cause smaller ion displacements, and accordingly less extra squeeze of $O^{2-}$ ions near the domain corners. Thus the twofold symmetry of ion displacement in the close $\textbf{+}$ and $\mathbf{\times}$ neighborhood, enabling extra $O^{-2}$ squeeze and Coulomb-oscillator localization, fades to a fourfold symmetry near the corners of the frame. The epaulets now have shrunk to localized Coulomb-oscillator islands about the $\textbf{+}$ and $\mathbf{\times}$ points, which destroys epaulet connectivity. This is illustrated in Fig. 5 by corner regions with lateral overswing of $3s$ electron oscillations between $nnn$ $O^{2-}$ ions, as it globally holds for the undoped compound.

For the overdoped regime in general, $x_{4} < x < 1/4$, the $Na^{+}$ pairs of the wider-spaced superlattice ${A^ \times }(x)$ are located at, or near some $\mathbf{\times}$ positions of the persisting $A^{+}(x_{4})$ domains. The superposition of ion displacements of twofold symmetry about the $Na^{+}$ pairs at the $\textbf{+}$ and $\mathbf{\times}$ positions, being perpendicular to each other, causes, as explained, a destruction of epaulet connectivity for \emph{those} $4a_{0} \times 4a_{0}$ domains of the persisting $A^{+}(x)$ superlattice that are \emph{affected} by $Na^{+}$ pairs at or near their $\mathbf{\times}$ positions. Conversely, all $A^{+}(x)$ domains that are \emph{not} occupied by $Na^{+}$ pairs at their $\mathbf{\times}$ points still maintain epaulet connectivity. It seems reasonable to model the frustration of $E^{+}$-$E^{+}$ epaulet connectivity by $E^{\times}$ epaulets in inverse 

\pagebreak
\includegraphics[width=5.05 in]{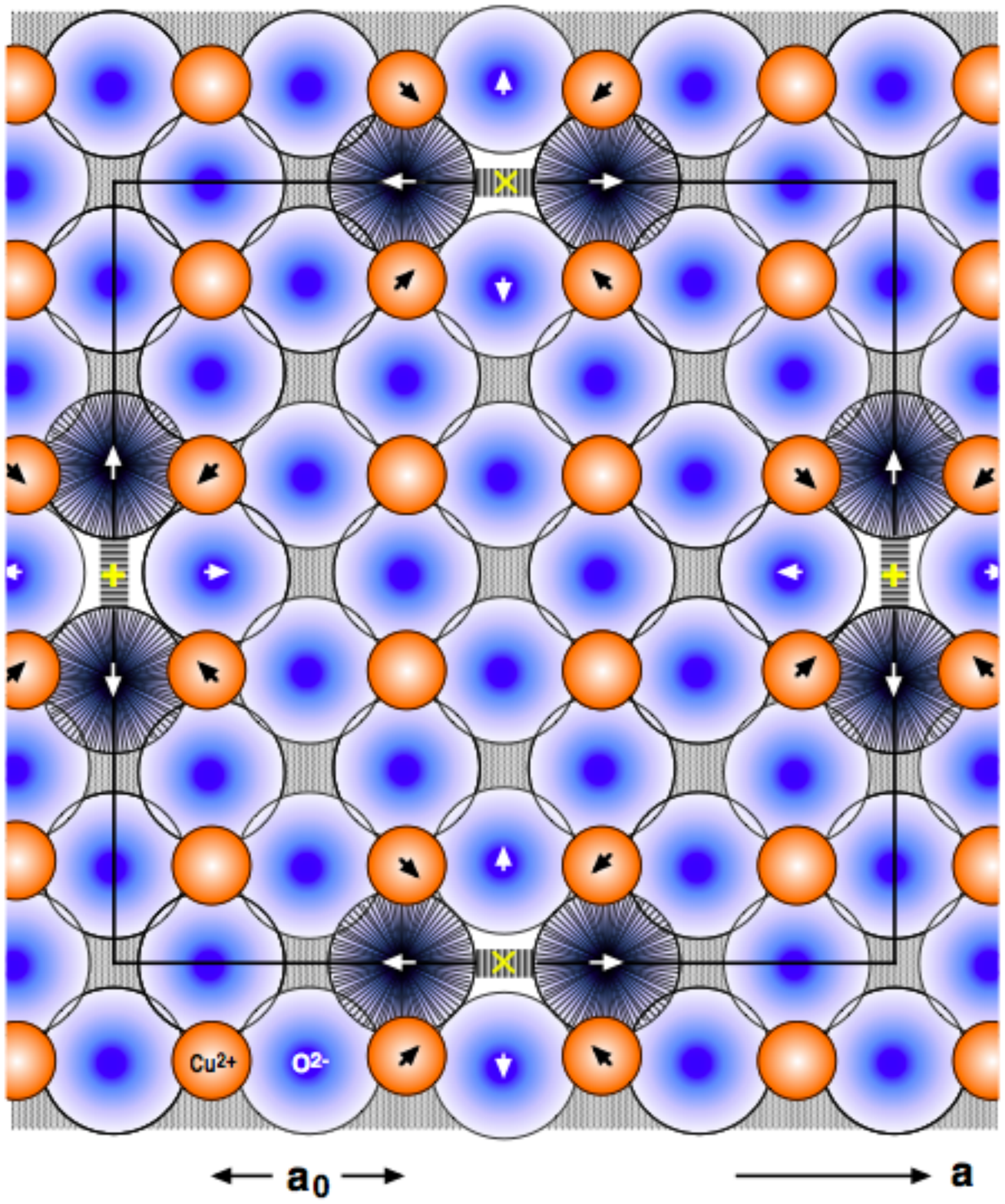}

\noindent FIG. 5. $CuO_{2}$ plane of $Ca_{2-x}Na_{x}CuO_{2}Cl_{2}$, $x = 0.25$, with $4{a_0} \times 4{a_0}$ domain.  Straight-line hatching indicates Coulomb oscillation of $3s$ electrons through oxygen nuclei and, laterally confined, across interionic regions. Shaded interionic space signifies lateral $3s$ overswing. Pairs of dopant $Na^{+}$, residing in the plane above and beneath, are located at \emph{both} the $\textbf{+}$ and $\mathbf{\times}$ positions on the domain frame. Partial compensation of electrostatic forces from the $\textbf{+}$ and $\mathbf{\times}$ centers give rise to weak ion displacement, and thus to \emph{insufficient} extra $O^{2-}$ squeeze, near the domain corners.  The corresponding $3s$ electrons are no longer prevented from lateral overswing, which destroys epaulet connectivity and accordingly superconductivity.

\pagebreak

\noindent relation to the spacing of the $A^{\times}(x)$ superlattice,
\begin{equation}
\vartheta (x) \equiv \frac{{4{a_0}}}{{{A^ \times }(\frac{3}{8} - x)}},  \; {x_4} < x \, \le \, 0.25,
\end{equation}
\noindent shown by the descending arc of the cusped dome in Fig. 4a.
 
If it was not for secondary influences, the doping dependence of the superconductive transition temperature, $T_{c}(x)$ would, by Eq. (3), have a cusp-shaped dome like the epaulet connectivity $\vartheta (x)$ in Fig. 4a.  However, experiment shows instead a smooth dome with a maximum near $x = 0.15$ and cessation of superconductivity near $x = 0.30$, as shown in Fig. 4b.  The reason for the dome's extension to the right could be that fluctuations enhance epaulet connectivity and impede epaulet-connection breaking by $Na^{+}$ pairs at $\mathbf{\times}$ positions more than the static situation would explain.

\section{SUPERCONDUCTIVITY IN $\mathbf{La_{2-x}Ae_{x}CuO_{4}, \; \; Ae = Ba, Sr}$}

The undoped compound $La_{2}CuO_{4}$ has two $CuO_{2}$ planes per crystal unit cell, each one sandwiched between two $LaO$ layers (see Fig. 6).  The $La^{3+}$ ions of each sandwiching layer reside above and beneath the empty center of the squares formed by the $Cu$-$O$ bonds---exactly as the $Ca^{2+}$ ions did in $Ca_{2}CuO_{2}Cl_{2}$. However, the two $LaO$-$CuO_{2}$-$LaO$ sandwiches are \emph{staggered} such that $La^{3+}$ ions of the top sandwich reside above the $O^{2-}$ ion at the center of the $La$-$La$ squares of the bottom sandwich. The staggering profoundly affects the epaulet connectivity $\vartheta (x)$, and by Eq. (3), the transition temperature of superconductivity in its dependence on doping, $T_{c}(x)$.

Undoped $La_{2}CuO_{4}$ is an insulator with an antiferromagnetic ground state.  When doped, vertical pairs of $Ae^{2+}$ ions (aligned parallel to the $c$ axis) substitute $La^{3+}$ ions in the top and bottom sandwiching layers of $La_{2-x}Ae_{x}CuO_{4}$, analogous to the $Na$ doping in $Ca_{2-x}Na_{x}CuO_{2}Cl_{2}$. The vertical $Ae^{2+}$ pairs in each the top and bottom sandwich form a square superlattice of spacing ${\overline{A}}^{+}(x)$ and ${\underline{A}}^{+}(x)$, respectively, given by Eq. (1). Here, and continuing below, an overline denotes a quantity of the top sandwich, and an underline one of the bottom sandwich.

It is helpful to imagine, for each sandwich, the vertical $Ae^{2+}$ pairs located at the $\textbf{+}$ marks at the middle of the left and right frame side, as in Fig. 3. The ion displacement in their neighborhood causes extra squeeze of some $O^{2-}$ ions and formation of $E^{+}$ epaulets. As in 

\pagebreak
\includegraphics[width=5.95in]{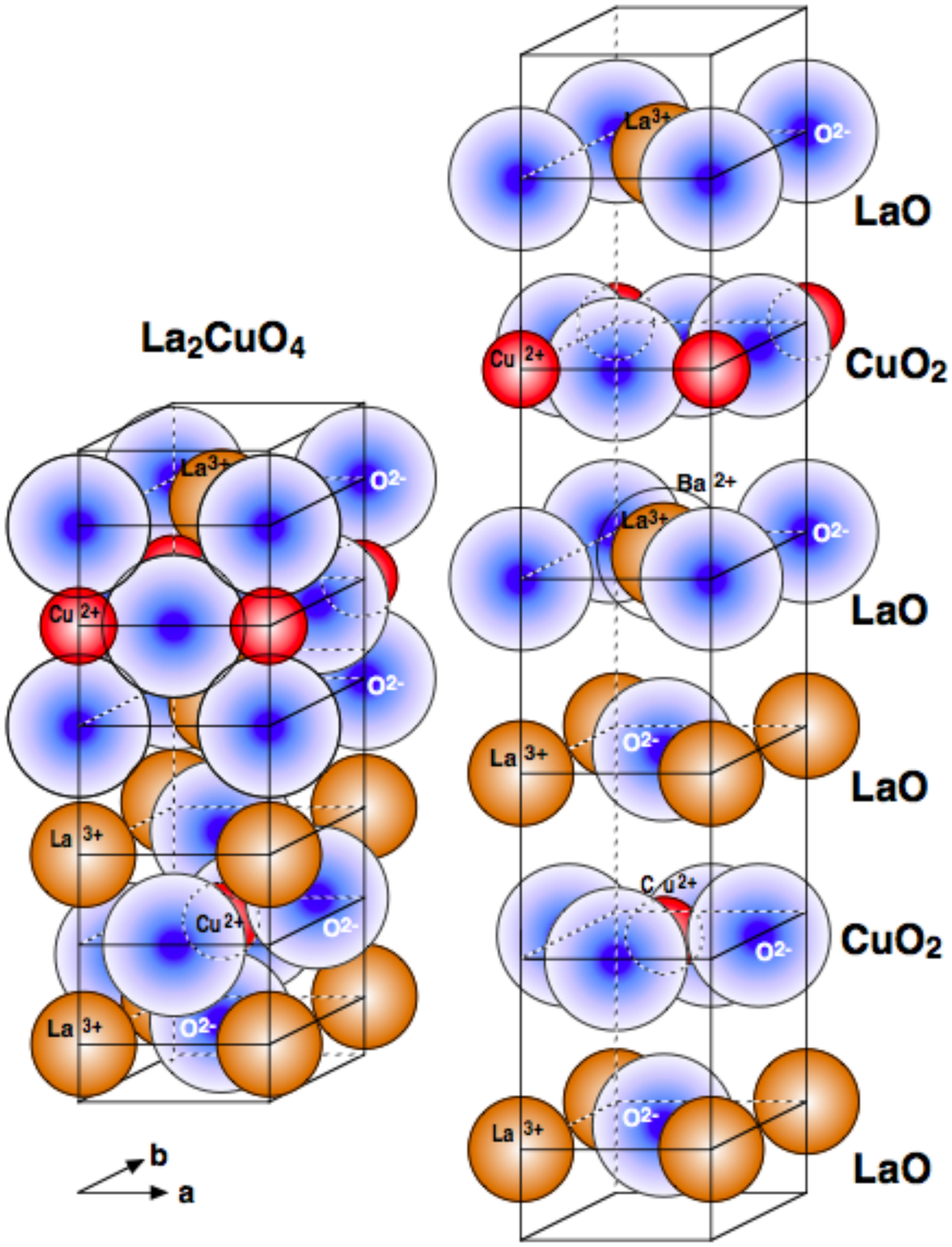}

\noindent FIG. 6.  Unit cell of $La_{2}CuO_{4}$ in hard-sphere display (left) and vertically exploded (right). The compound has a staggered two-sandwich structure, where each ``sandwich'' consists of a $CuO_{2}$ plane between two $LaO$ layers.  When doped, $La^{3+}$ ions (of radius $r=1.06$ {\AA}) are substituted by $Ba^{2+}$ ($r = 1.34$ {\AA}, indicated by the circle in the third layer on the right) or by $Sr^{2+}$ ($1.12$ {\AA}).
\pagebreak

\noindent the previous compound with a single $CuO_{2}$ plane per unit cell, antiferromagnetism is destroyed at $x = x_{10} \equiv 0.02$. Likewise, in each $CuO_{2}$ plane superconductivity emerges at $x = x_{6}$ with the onset of $E^{+}$-$E^{+}$ epaulet overlap and it ceases at $x = 1/4$ by $E^{\times}$ frustration as discussed above. Accordingly the curve of epaulet-connectivity $\vartheta (x)$ in Fig. 7a for $La_{2-x}Ba_{x}CuO_{4}$ (solid line) merges with $\vartheta (x)$ for $Ca_{2-x}Na_{x}CuO_{2}Cl_{2}$ (dashed line) near $x_{6}$ and for $x \to 1/4$. However, a new aspect arises in the present double-sandwich situation. The ensuing mechanism will plough into the previous cusped $\vartheta (x)$ dome (dashed line) a deep furrow at $x_{4}$, leaving a double hump (solid line in Fig. 7a). 

Let's assume that in the relevant underdoped regime, $x_{10} < x < x_{4}$, dopant $Ae^{2+}$ pairs occupy the $\textbf{+}$ positions at the left and right side of their respective domain frame, that is, ${\overline{Ae}}^{2+}$ at $\overline{\textbf{+}}$, to be called ``plus-top,'' and $\underline{Ae}^{2+}$ at $\underline{\textbf{+}}$ (``plus-bottom''). Minimization of electrostatic repulsion places the $\underline{Ae}^{2+}$ pairs of the bottom sandwich below the midpoint of the upper and lower domain-frame bar of the top sandwich, denoted $\overline{\mathbf{\times}}$ (``cross-top''). Thus the $\underline{Ae}^{2+}$ pairs of the bottom sandwich, both of the unit cell under consideration and of the unit cell above (along $+c$), are at the position where they frustrate epaulet connectivity in the top $CuO_{2}$ plane, $\underline{\textbf{+}} = \overline{\mathbf{\times}}$ (consult Fig. 5). 

The peculiarity of superconductivity in $La_{2-x}Ae_{x}CuO_{4}$ arises from the ion displacements in the $CuO_{2}$ plane of one sandwich due to the electrostatic force from the staggered $Ae^{2+}$ pairs of the other sandwich, both above and beneath. It gives rise to counteracting effects, here explained for the top sandwich (and correspondingly valid for the bottom sandwich): Increased doping in the superconducting underdoped regime, $x_{6} < x < x_{4}$, tends to enhance, in the top $CuO_{2}$ plane, epaulet connectivity by ${\overline{E}}^{+}$-${\overline{E}}^{+}$ overlap, as mentioned.  But it also tends to frustrate those ${\overline{E}}^{+}$-${\overline{E}}^{+}$ connections by the perpendicular ion displacements about the $\overline{\mathbf{\times}}$ positions in the top $CuO_{2}$ plane due to the $\underline{Ae}^{2+}$ pairs at $\underline{\textbf{+}}$ in the bottom sandwich of the same cell and of the unit cell above.

Recall the single $CuO_{2}$-plane compound $Ca_{2-x}Na_{x}CuO_{2}Cl_{2}$, where we had epaulet connectivity \emph{exclusively} by $E^{+}$-$E^{+}$ overlap in the underdoped regime, followed by $E^{\times}$ frustration in the overdoped regime, both of which resulted in the cusped $\vartheta (x)$ dome of Fig. 5a.  In contrast, ${\overline{E}}^{+}$-${\overline{E}}^{+}$ overlap and frustration due to ${\underline{E}}^{+}$ presence (under and over the $\overline{\mathbf{\times}}$ positions in the top $CuO_{2}$ plane), occur now \emph{simultaneously} in the underdoped regime of the staggered double $CuO_{2}$-plane compound, $La_{2-x}Ba_{x}CuO_{4}$. This causes the first hump in Fig. 8a.  Quantitatively, it is modeled by

\pagebreak
\includegraphics[width=6in]{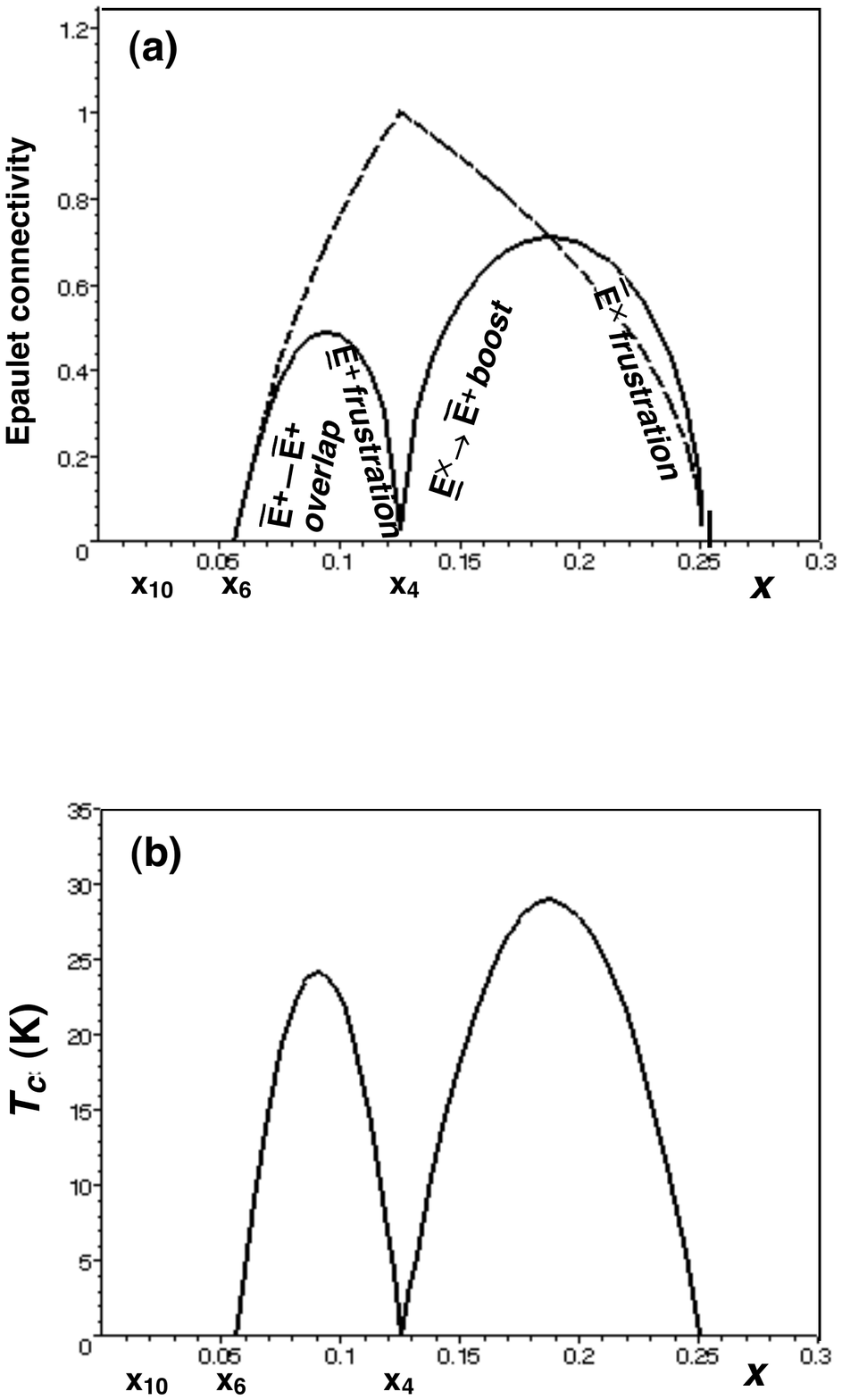}

\noindent FIG. 7. (a) Double-humped doping dependence of epaulet connectivity $\vartheta (x)$ for the double $CuO_{2}$-plane compound $La_{2-x}Ba_{x}CuO_{4}$.  The cusped dome (dashed line) is the same as in Fig. 5a for a single $CuO_{2}$ plane. (b) Corresponding transition temperature of superconductivity $T_{c}(x)$ by Eq. (3), causing complete suppression of superconductivity at $x = x_{4} = 1/8$.

\pagebreak

\begin{equation}
\vartheta (x) = \frac{{L_{E} - {A^ + }(x)}}{{2{a_0}}} \; \frac{{4\sqrt 2 {a_0}}}{{{A^ \times }(\frac{1}{4} - x)}} \; , \; {x_6} \, \le \, x \, \le \; {x_4}.
\end{equation} 
\noindent The first term in Eq. (6) arises from ${\overline{E}}^{+}$-${\overline{E}}^{+}$ overlap. The second term expresses a modulation due to simulataneous frustration from ${\underline{E}}^{+}$. Note that the second (frustrating) term has the same doping dependence as in Eq. (5) except for a shift of the disappearing end of the descending arc from $x=1/4$ to $1/8$.

As in the $Ca_{2-x}Na_{x}CuO_{2}Cl_{2}$ case, the $\overline{A}^{+}(x_{4})$ superlattice, containing $\overline{Ae}^{2+}$ pairs at the $\overline{\textbf{+}}$ positions, keeps persisting in the overdoped regime. Likewise a new superlattice, $\overline{A}^{\times}(x)$ and $\underline{A}^{\times}(x)$, now evolves in the top and bottom $CuO_{2}$ plane, respectively, with spacing giving by Eq. (4). Because of sandwich staggering, some $\underline{Ae}^{2+}$ pairs at $\underline{\mathbf{\times}}$ positions of the bottom sandwich align (or nearly align) with  $\overline{\textbf{+}}$ positions of the top sandwich. In these instances the neighbor ions in the top $CuO_{2}$ plane are subject to the electrostatic force from both the $\overline{Ae}^{2+}(\overline{\textbf{+}})$ pair in the top sandwich and the aligned $\underline{Ae}^{2+}(\underline{\mathbf{\times}})$ pair in the bottom sandwich of the same cell and of the unit cell above. This gives a \emph{boost} to the $\overline{E}^{+}$ epaulet, overcoming the impasse at the corners of the $4a_{0} \times 4a_{0}$ domain caused by $\overline{E}^{\times}$ epaulets due to $\overline{Ae}^{2+}(\overline{\mathbf{\times}}$) pairs in the top sandwich. Thus with increasing doping in the overdoped regime this $\underline{E}^{\times} \to \overline{E}^{+}$ boost establishes a new $\overline{E}^{+}$-$\overline{E}^{+}$ epaulet connectivity in inverse relation to the shrinking of the $\underline{A}^{\times}(x)$ superlattice. The $\underline{E}^{\times}$ $\to$ $\overline{E}^{+}$ boost gives rise to the ascending arc of the second hump in Fig. 8a. The subsequent descending arc originates, as mentioned, with the filling of $\overline{Ae}^{2+}(\overline{\mathbf{\times}}$) pairs in the top $CuO_{2}$ plane and their frustrating effect on $\overline{E}^{+}$-$\overline{E}^{+}$ epaulet connectivity. Quantitatively, the combined effect of simultaneous $\underline{E}^{\times}$ $\to$ $\overline{E}^{+}$ boost and $\overline{E}^{\times}$ frustration on epaulet connectivity in the overdoped regime is modeled by
\begin{equation}
\vartheta (x) = \frac{{4\sqrt 2 {a_0}}}{{{A^ \times }(x)}} \;
\frac{{4{a_0}}}{{{A^ \times }(\frac{3}{8} - x)}} \; ,  \; {x_4} < x \, \le \, 0.25.
\end{equation}
\noindent Note that the first term in Eq. (7), expressing the  $\underline{E}^{\times} \to \overline{E}^{+}$ boost and modulating the $\overline{E}^{\times}$ frustration of the second term, is mirror-symmetric (with respect to $x = 1/8$) to the second term in Eq. (6) that there modulated $\overline{E}^{+}$-$\overline{E}^{+}$ connectivity by frustration from $\underline{Ae}^{2+}($\underline{\textbf{+}}$)$ pairs aligned with $\overline{\mathbf{\times}}$ positions.


Combining epaulet connectivity $\vartheta(x)$ with a compound-specific robustness factor $\rho$, Eq. (3), accounts for the double-humped doping dependence of the superconducting transition temperature $T_{c}(x)$ of $La_{2-x}Ba_{x}CuO_{4}$ shown in Fig. 7b. Experimentally it is found that 

\pagebreak
\includegraphics[width=6in]{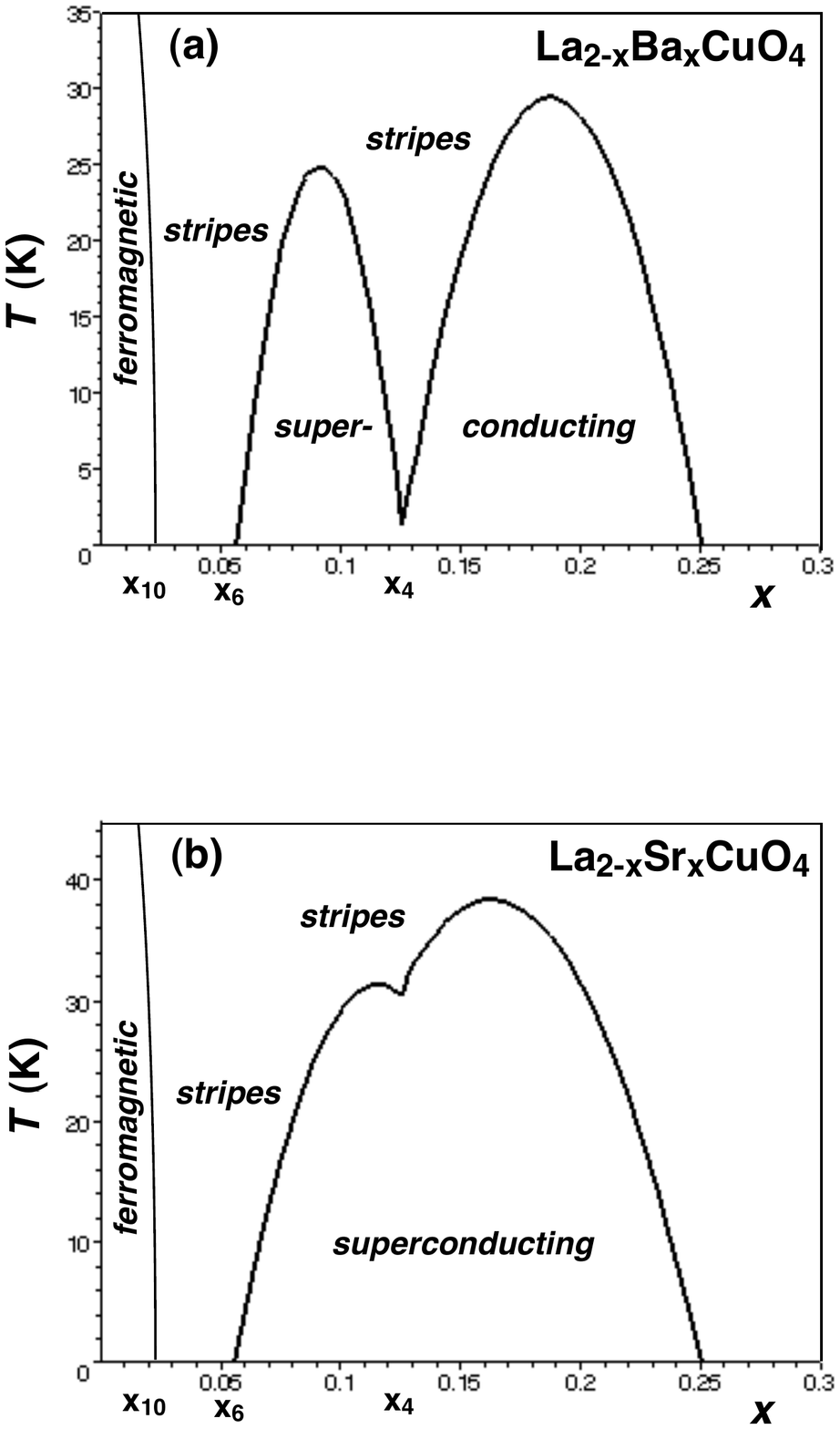}

\noindent FIG. 8. Phase diagram of (a) $La_{2-x}Ba_{x}CuO_{4}$ and (b) $La_{2-x}Sr_{x}CuO_{4}$ as experimentally observed. Note the different temperature scales.
\pagebreak

\noindent superconductivity is rarely completely suppressed in $La_{1.875}Ba_{0.125}CuO_{4}$ to $T_{c}(x_{4}) = 0$, as it at first appeared,\cite{5} but most frequently to $T_{c}(x_{4}) = 2.5$ K, or 4 K, or even 10 K, as summarily depicted in Fig. 8a.\cite{6,7,8,9}  Apart from crystal quality, the reason is that at doping $x = x_{4}$ the frustration of $\overline{E}^{+}$-$\overline{E}^{+}$ connectivity in the top $CuO_{2}$ plane by the more distant $\underline{Ae}^{2+}$ pairs at $\underline{\textbf{+}}$ positions in the bottom sandwich of the same and the above unit cell is less effective than their frustration by the closer $\overline{Ae}^{2+}$ pairs at $\overline{\mathbf{\times}}$ positions in the top sandwich at doping $x = 0.25$.

In contrast, the transition temperature $T_{c}(x)$ of the related compound $La_{2-x}Sr_{x}CuO_{4}$ possesses no deep cleavage of a $T_{c}(x)$ dome but merely a small dip at $x = x_{4}$, resulting in a kinked dome (see Fig. 8b). The distinctly different $T_{c}(x)$ profiles of the sister compounds $La_{2-x}Ba_{x}CuO_{4}$ and $La_{2-x}Sr_{x}CuO_{4}$ are a consequence of different phases of the crystal structure in combination with different cation sizes.  

Undoped $La_{2}CuO_{4}$ has an orthorhombic crystal structure with lattice constants $b_{0} > a_{0}$.  This is cohesively advantageous as it permits closer proximity of the top and bottom sandwich at adjacent $LaO$ layers (layers 3 and 4 in Fig. 6). In a simplified view, the bottom sandwich is slightly displaced in the, say, negative $b$ direction, and the top sandwich slightly in the positive $b$ direction. (More accurately, $CuO_{6}$ octahedra in both sandwiches undergo a tilt.) This enables a tighter settling of the ions in the $3^{rd}$ and $4^{th}$ layers by their extension into the the available space in the $a_{0} \times b_{0}$ rectangle of the opposite layer.  Replacement of $La^{3+}$ ions ($r = 1.06$ {\AA}) with larger $Ba^{2+}$ dopant ions ($r = 1.34$ {\AA}, indicated in the $3^{rd}$ layer of Fig. 6) widens the inter-sandwich spacing, with consequent increase of the $c_{0}$ lattice constant and decrease of $b_{0}$.  The $a_{0}$ lattice parameter remains essentially constant throughout the $Ae$ doping regime.

When, with increasing doping, $b_{0} = a_{0}$ is reached, a structural transition from the low-temperature orthorhombic (LTO) phase to a low-temperature tetragonal (LTT) phase occurs.  This happens for $x = 0.1$ in $La_{2-x}Ba_{x}CuO_{4}$ and at a larger doping ratio, $x = 0.2$, in $La_{2-x}Sr_{x}CuO_{4}$ due to the smaller size of $Sr^{2+}$ ($r = 1.12$ {\AA}) than $Ba^{2+}$.\cite{10,11} The kinked $T_{c}(x)$ dome of $La_{2-x}Ba_{x}CuO_{4}$ in Fig. 8b can be regarded as the joining of smeared-out double humps due to rectangular instead of square domains, induced by dopant $Sr^{2+}$ pairs (orthorhombic distortion).  This also extends the cessation of superconductivity in $La_{2-x}Ba_{x}CuO_{4}$ to a doping ratio somewhat beyond $x = 0.25$ (see Fig. 8b).

When $La_{1.875}Ba_{0.125}CuO_{4}$ was compressed with hydrostatic pressure $p$ from 0 to 2.2 GPa, 

\pagebreak
\includegraphics[width=5.95in]{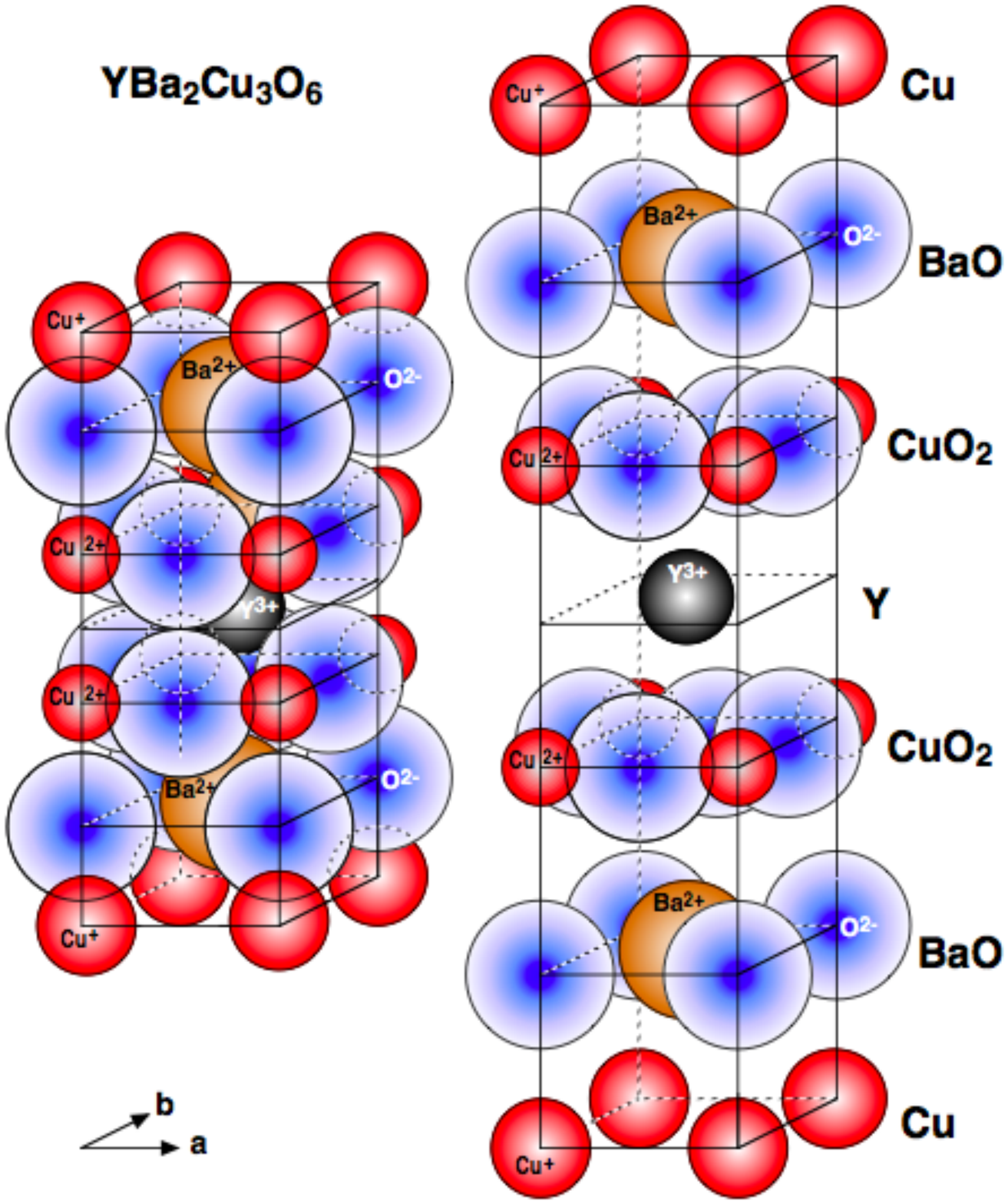}

\noindent FIG. 9.  Unit cell of $YBa_{2}Cu_{3}O_{6}$ in hard-sphere display (left) and vertically exploded (right).
\pagebreak

\includegraphics[width=5.95in]{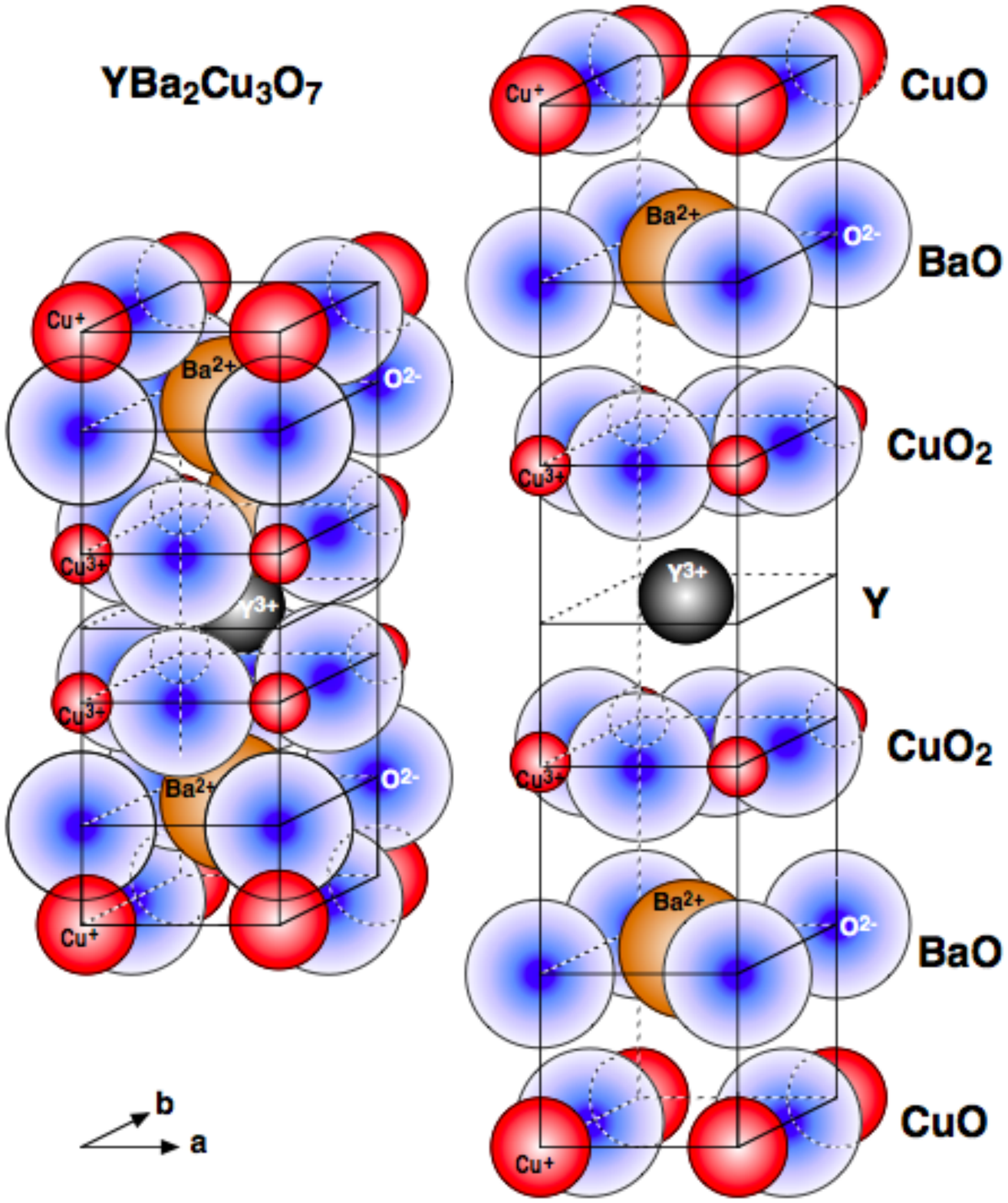}

\noindent FIG. 10.  Unit cell of $YBa_{2}Cu_{3}O_{7}$ in hard-sphere display (left) and vertically exploded (right).
\pagebreak

\noindent the suppressed transition temperature at the double-hump furrow, $T_{c}(x_{4})$, steadily increased from 9.6 K to 12.2 K.\cite{9} Qualitatively, an increase of superconductivity with pressure can be expected on general grounds (squeeze effect) and possibly also by distortion of the LTT phase under external pressure (reversing the influence of large $Ae^{2+}$ dopant ions).

\section{SUPERCONDUCTIVITY IN $\mathbf{YBa_{2}Cu_{3}O_{6+x}}$}

Contrary to the copper oxide families with cation doping, discussed so far, $YBa_{2}Cu_{3}O_{6+x}$ involves oxygen variation. It contains $CuO_{2}$ planes---two per unit cell, like the previous sister compounds, but now aligned---that are susceptible to superconductivity by confinement of lateral $3s$ electron oscillation due to displacement-induced extra squeeze of $O^{2-}$ ions. While for the cation-doped cuprates ion displacements in the $CuO_{2}$ plane are caused by electrostatic force from pairs of dopant ions in the sandwiching layers, in the case of oxygen-enriched $YBa_{2}Cu_{3}O_{6+x}$ the displacing electrostatic force emanates from the $CuO_{2}$ plane itself---from \emph{triple} ionized copper atoms, $Cu^{3+}$.

Figures 9 and 10 show a unit cell of $YBa_{2}Cu_{3}O_{6+x}$ for the extreme cases with $x=0$ and $x=1$. Simple stoichiometry confirms the assigned ion charges. The compound with most oxygen depletion, $YBa_{2}Cu_{3}O_{6}$, is an insulator with tetragonal crystal structure, $a_{0}=b_{0}$. The two $CuO_{2}$ planes are separated by a layer of $Y^{3+}$ ions and are covered on the opposite sides by $BaO$ layers. At the very top and bottom of the unit cell is each a copper layer, ionized to $Cu^{+}$.

With increasing oxygen content, $O$ atoms settle between the $Cu^{+}$ ions in the terminal layers (compare Figs. 9 and 10). Each such $O$ atom in the top layer deprives a \emph{pair} of $Cu^{2+}$ ions in the upper $CuO_{2}$ plane of \emph{one} electron (and likewise deprives of one electron a $Cu^{2+}$ pair in the lower $CuO_{2}$ plane of the unit cell \emph{above}, along $+c$). This results in an $O^{2-}$ ion and a $Cu^{2+}$-$Cu^{3+}$ pair in those places. (The equivalent holds for an $O$ atom in the bottom terminal layer of the unit cell under consideration and its flanking $CuO_{2}$ planes.) The
fact that the cation charge is \emph{different} at copper sites bracketing $O^{2-}$ ions in the $CuO_{2}$ plane beneath \emph{filled} or \emph{empty} $O^{2-}$ sites in the top terminal layer has been shown with nuclear magnetic resonance (NMR) measurements.\cite{12}

The $Cu^{2+}$-$Cu^{3+}$ pair in the $CuO_{2}$ plane will ``fight'' for the single $3d^{9}$ electron between them, causing a fluctuation of the electron. On average each copper ion of the pair possesses 

\pagebreak

\includegraphics[width=6.2in]{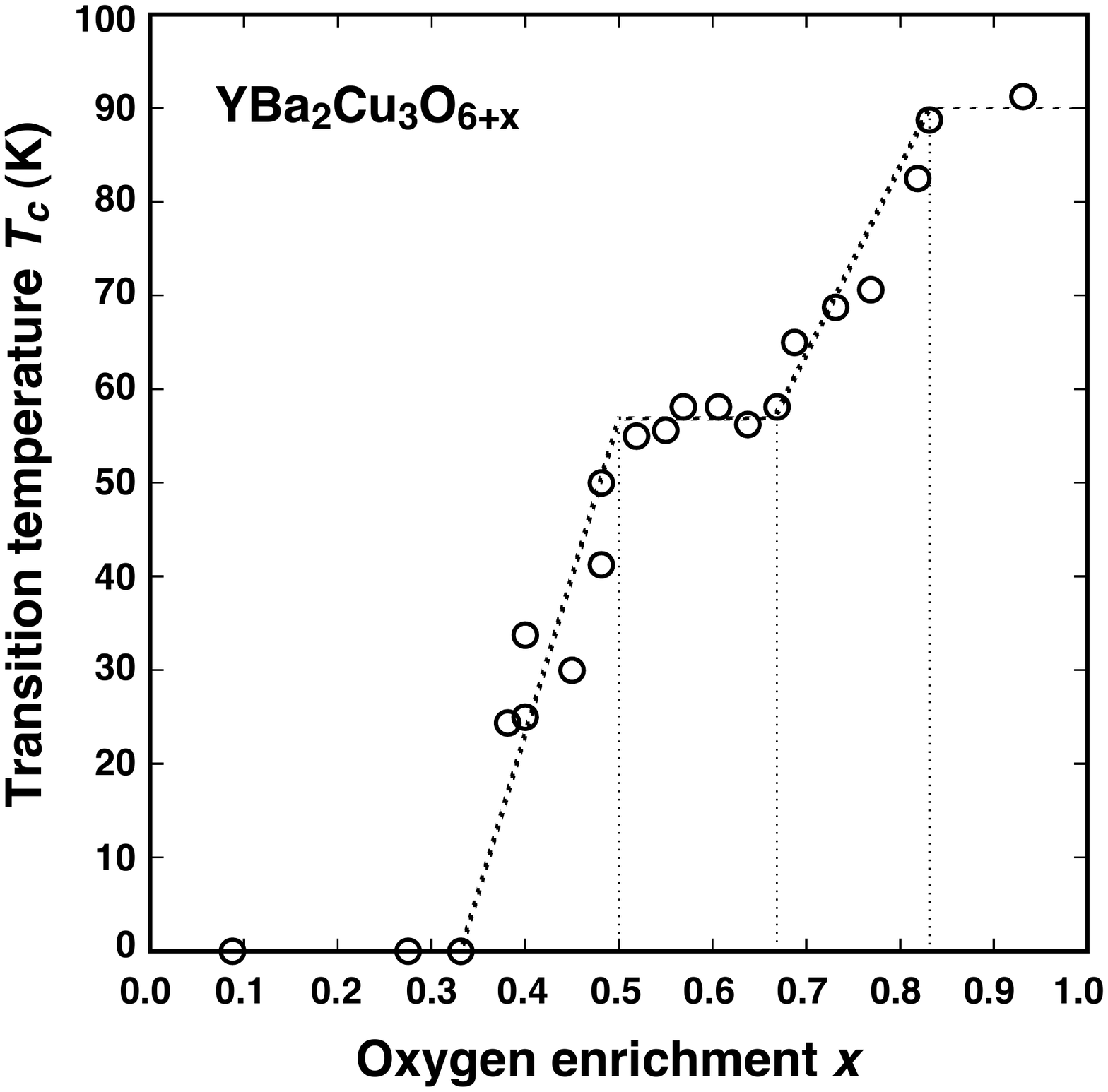}

\noindent FIG. 11. Experimental values of the superconducting transition temperature $T_{c}$ of $YBa_{2}Cu_{3}O_{6+x}$ in dependence of oxygen enrichment $x$. Hatched lines are guides to the eye. The $x$ range is divided by values $x_{n} = n/6$, $n = 2, ...,6$, marked by vertical lines, into ramp and plateau segments. Data taken from Fig. 3 of Ref. 13.
\pagebreak

\includegraphics[width=7in]{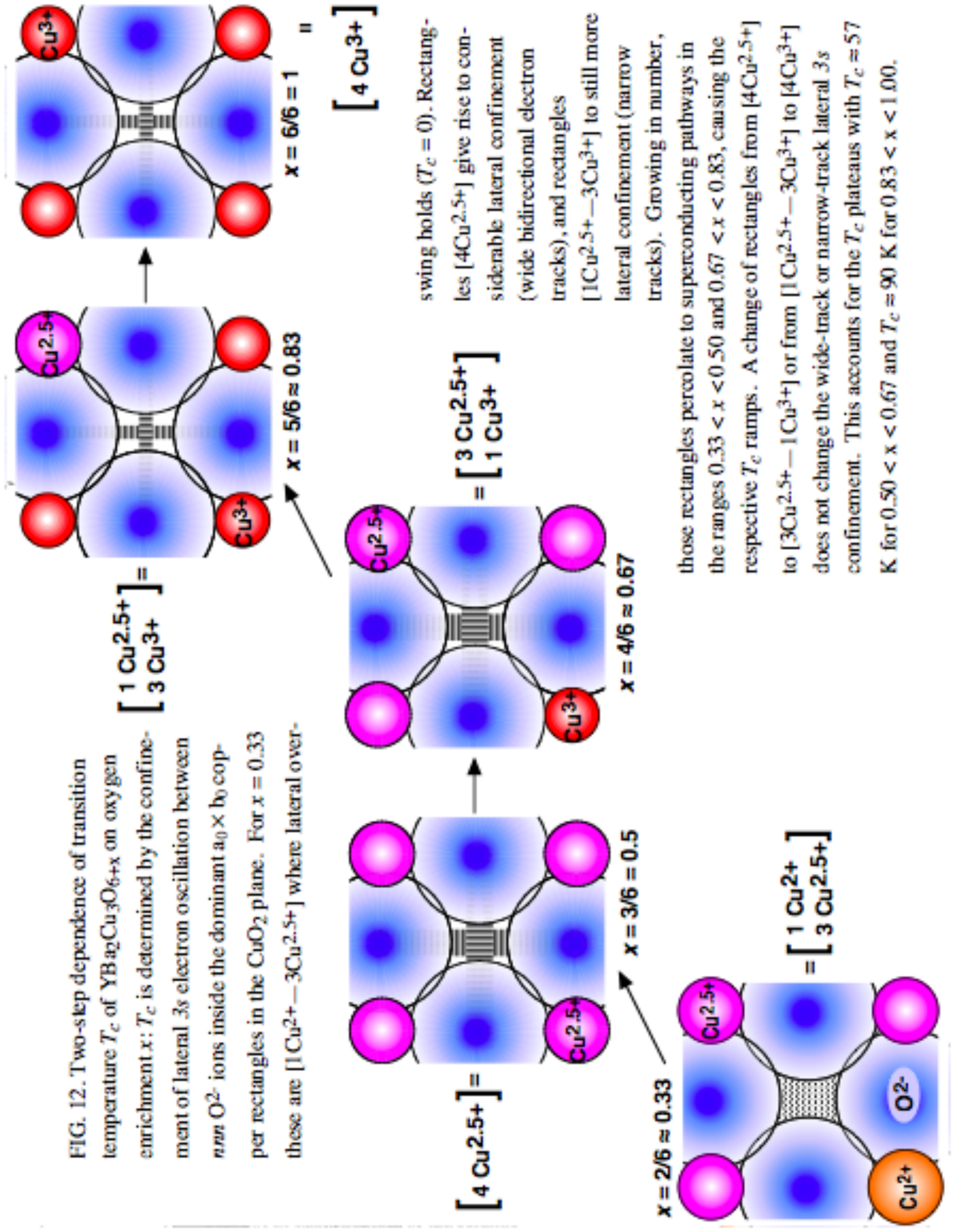}

\pagebreak

\noindent the $3d^{9}$ electron half the time. Thus the \emph{time-average} of the fluctuating $Cu^{2}$-$Cu^{3+}$ pair will be denoted as a $Cu^{2.5+}$ pair. As we'll see, consideration of time-average $Cu^{2.5+}$ pairs is sufficient to explain the dependence of superconductivity on oxygen enrichment $x$.

The lattice constant $c_{0}$ \emph{decreases} ($11.82$ {\AA} $\to 11.78$ {\AA}) with increasing oxygen content ($x=0 \to 0.33$) because of enhanced electrostatic attraction that arises from the new $O^{2-}$ and $Cu^{3+}$ ions. The oxygen enrichment thus causes \emph{additional} compression of the $CuO_{2}$ planes (``extra'' squeeze). The other lattice parameter, in contrast, stays essentially constant under oxygen increment (${a_0} \simeq 3.86$ {\AA}) due to much cancellation of the $CuO_{2}$ plane's outward bulging from compression along the $c$ axis and enhanced attraction between  $Cu^{3+}$ and $O^{2-}$ ions within the plane.

When oxygen enrichment has reached $x = 1/3$ a phase transition occurs. It causes a resettlement of all enriching $O$ atoms (ionized to $O^{2-}$) parallel to the crystal's $b$ axis. The new arrangement of the $O^{2-}$ ions in the terminal layers gives rise to so-called ``oxygen chains'' (see Fig. 10). The crystal then assumes orthorhombic symmetry, $b_{0} > a_{0}$, where for the oxygen-enriched regime $x = 0.33 \to 1$ the lattice constants $b_{0}$ and $a_{0}$ symmetrically bifurcate with parabolic tongs about the constant quasi-tetragonal value $\overline{a}_{0} \equiv  (a_{0} + b_{0})/2 \simeq 3.86$ {\AA} to $a_{0} \simeq 3.82$ {\AA} and $b_{0} \simeq 3.89$ {\AA}. The other lattice constant keeps decreasing in that $x$-range from $c_{0} = 11.78$ {\AA} $\to 11.68$ {\AA}.\cite{13} 

Superconductivity is absent when the crystal is tetragonal ($0 < x < 1/3$) but emerges at the transition to the orthorhombic phase at $x = 1/3$. The profile of the transition temperature $T_{c}(x)$ differs distinctly from the dome or humps of the previous compounds: (1) It \emph{never decreases} with oxygen enrichment, rising from $T_{c}(0.33)=0 \to T_{c}(0.93) = 92$ K (no measurements are reported for $0.93 < x < 1$). (2) Instead of a smooth semidome arc, the increase of $T_{c}(x)$ occurs essentially in \emph{two steps,} each with a sharp rise followed by a range of constant value, called ``plateau,'' to wit $T_{c} \simeq 57$ K and $T_{c} \simeq 90$ K (see Fig. 11).\cite{13}  The width $\Delta x$ of the ramps and plateaus is essentially equal, dividing the orthorhombic phase of oxygen enrichment into four parts, $\Delta x = \frac{2}{3}/4 = 1/6 \simeq 0.17$.
 
As outlined in the overview of Fig. 12, the two-step increase of $T_{c}(x)$ is determined by the lateral oscillation of $3s$ electrons between $nnn$ $O^{2-}$ ions in the $CuO_{2}$ plane, characterized by the four copper ions that corner each each set of two perpendicular $O^{2-}$ pairs. Generally speaking, extra squeeze of the $O^{2-}$ pairs enclosed in an $a_{0} \times b_{0}$ copper rectangle increases with copper ionization at the corners. Specifically, the squeeze of $O^{2-}$ ions in the copper 

\pagebreak
\includegraphics[width=5.05in]{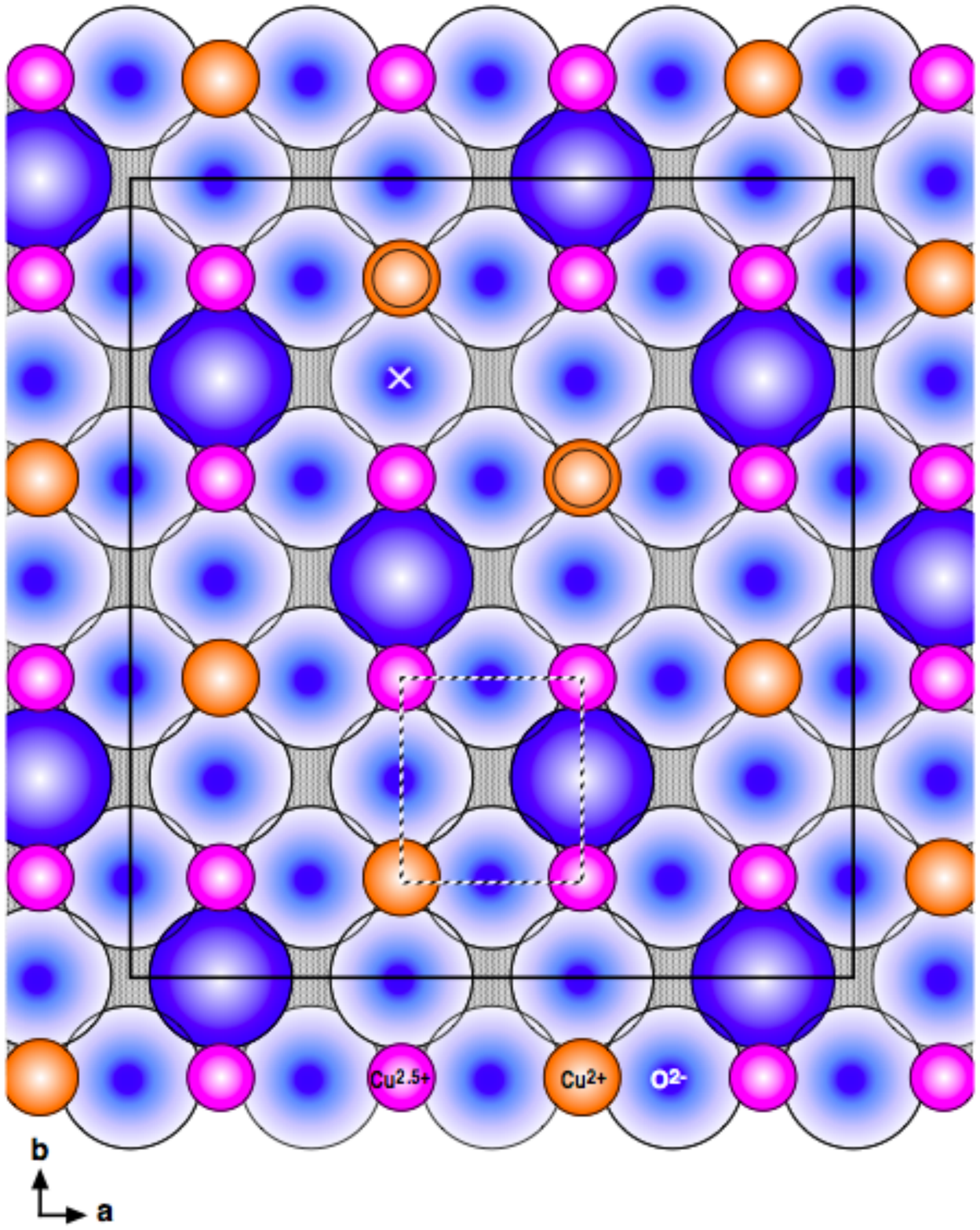}

\noindent FIG. 13.  Upper $CuO_{2}$ plane in $YBa_{2}Cu_{3}O_{6.33}$: The solid frame delimits a $4{a_0} \times 4{b_0}$ domain of (exaggerated) orthorhombic structure. Each $O^{2-}$ ion with bold shading sits beneath an $O^{2-}$ ion in the oxygen chain above (along $+c$). In the $CuO_{2}$ plane those $O^{2-}$ ions are bracketed each by a pair of copper ions $Cu^{2.5+}$ of smaller size and nominal charge ``$+2.5$'', representing the time-average of a fluctuating $Cu^{2+}$-$Cu^{3+}$ pair. The $CuO_{2}$ plane is paved by ${a_0} \times {b_0}$ copper rectangles, each cornering two perpendicular $nnn$ $O^{2-}$ ions. The ionization of a [$1Cu^{2+}$---$3Cu^{2.5+}$] rectangle, marked by hatched lines, and the ensuing ion displacement are \emph{insufficient} to prevent lateral overswing of $3s$ electrons between $nnn$ $O^{2-}$ ions. (The $O^{2-}$ with $\times$ mark and two nearby $Cu^{2+}$ with a ring prepare for changes in Fig. 14.)
\pagebreak

\includegraphics[width=5.05in]{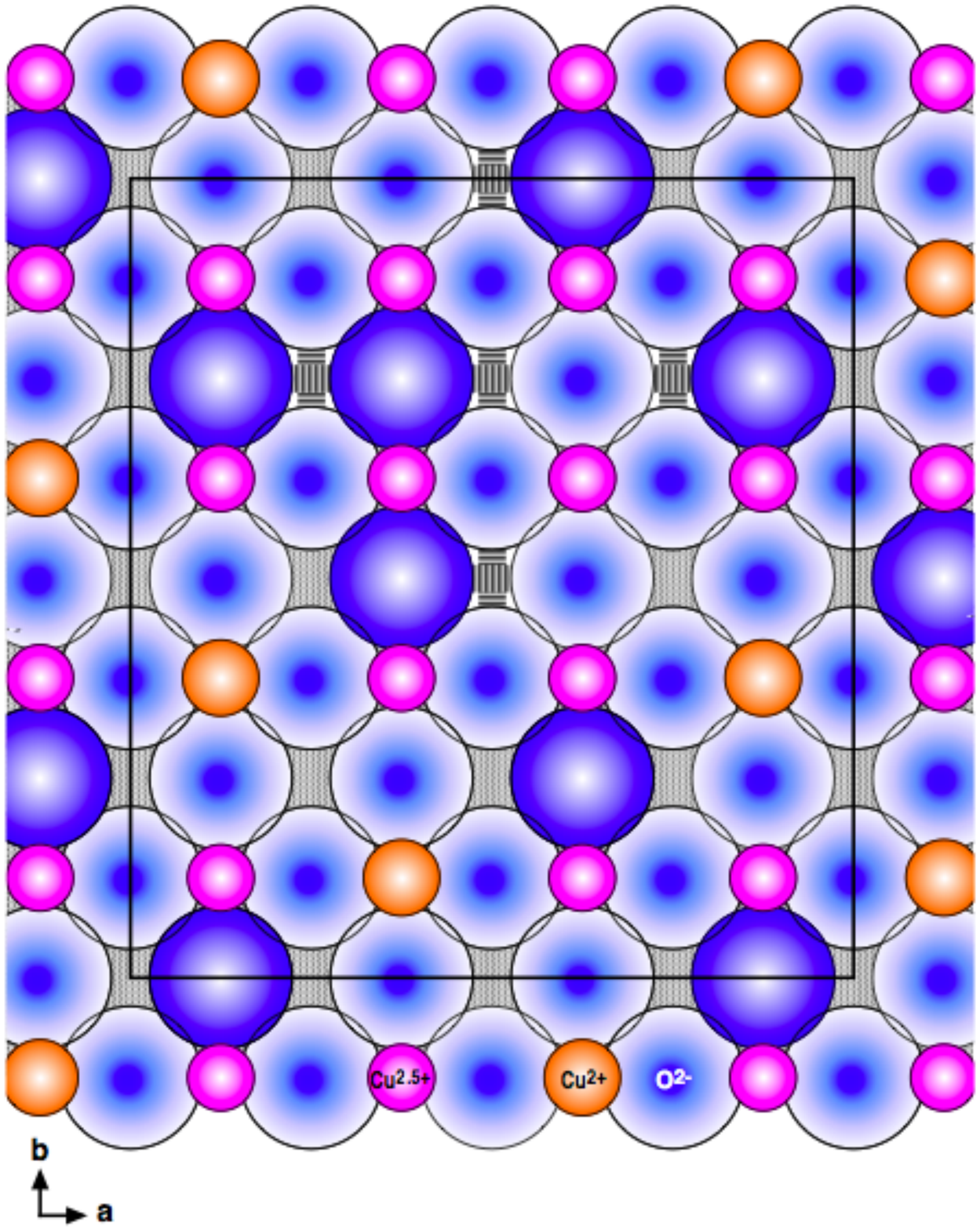}

\noindent FIG. 14.  Upper $CuO_{2}$ plane in $YBa_{2}Cu_{3}O_{6+x}$ for $x = 1/3$ plus \emph{one} additional $O$ atom in the oxygen chain above the $\times$ mark in Fig. 13. Its ionization to $O^{2-}$ converts two $Cu^{2+}$ ions, marked by rings in Fig. 13, to $Cu^{2.5+}$. This gives rise to several [$4Cu^{2.5+}$] rectangles with confined $3s$ electron oscillation between $nnn$ $O^{2-}$ ions (depicted by wide electron tracks) that combine to a cross-shaped $3a_{0} \times 3b_{0}$ superconducting island.

\pagebreak

\noindent rectangles [$4Cu^{2+}$] in the $CuO_{2}$ planes of non-enriched $YBa_{2}Cu_{3}O_{6}$, and the \emph{extra} squeeze in the rectangles [$3Cu^{2+}$---$1Cu^{2.5+}$], [$2Cu^{2+}$---$2Cu^{2.5+}$] and [$1Cu^{2+}$---$3Cu^{2.5+}$] that occur in the tetragonal, oxygen-enriched compound, $0 < x < 0.33$, is \emph{insufficient} to prevent lateral overswing, resulting in a continuation of $T_{c} = 0$. However, \emph{more} ionization of copper rectangles, to [$4Cu^{2.5+}$] and [$3Cu^{2.5+}$---$1Cu^{3+}$], causes enough extra squeeze of $O^{2-}$ ions to achieve lateral confinement of $3s$ electron oscillation between $nnn$ $O^{2-}$ ions (wide bidirectional electron tracks) and thereby the emergence of superconductivity. Still more ionized copper rectangles, [$1Cu^{2.5+}$---$3Cu^{3+}$] and [$4Cu^{3+}$], cause more lateral $3s$ confinement to narrow electron tracks.

Superconductivity is destroyed when the lateral $3s$ confinement becomes subject to dynamic ion displacement from thermal motion (lattice vibration of relevant frequency $\omega$) with enough phonon energy, 
$\varepsilon  = \hbar \omega  > {k_B}T_{c}$, to cause lateral overswing. In the oxygen-enrichment ranges $0.5 < x < 0.67$ and $0.83 < x < 1$, where copper rectangles with wide-track and, respectively, narrow-track electron paths pave the entire $CuO_{2}$ plane, this results in the plateaus of $T_{c} \simeq 57$ K and $T_{c} \simeq 90$ K. Lower transition temperature $T_{c}$ is necessary to destroy superconductivity in the ramp regions of oxygen enrichment because \emph{fewer} copper rectangles with confined electron tracks are involved in the percolation network of superconducting paths.

Equipped with Fig. 12 as an orientation device, we can now explore the sitation in greater detail. To this end we inspect the $CuO_{2}$ plane at the values of oxygen enrichment, $x_{n} = n/6, \; n = 2, ..., 5$, that mark the onset of a $T_{c}(x)$ ramp or plateau. Also, we'll observe the ensuing changes when these cases are enriched by \emph{one} additional $O$ atom in the oxygen chain. Figure 13 shows the upper $CuO_{2}$ plane of $YBa_{2}Cu_{3}O_{6+x}$ in the orthorhombic phase with oxygen enrichment $x = 1/3$ that marks the onset of superconductivity. Boldly shaded $O^{2-}$ ions in that plane reside beneath enriching $O^{2-}$ ions in the oxygen chains of the top terminal layer. They are flanked by fluctuating $Cu^{2}$-$Cu^{3+}$ pairs, here depicted as time-averaged $Cu^{2.5+}$ pairs, slightly smaller in size than the $Cu^{2+}$ ions otherwise. An inspection of the $CuO_{2}$ plane shows that it is paved by a $\frac{1}{3}$-minority of [$2Cu^{2+}$---$2Cu^{2.5+}$] and a $\frac{2}{3}$-majority of [$1Cu^{2+}$---$3Cu^{2.5+}$] copper rectangles. Their cation charges and the resulting ion displacements are \emph{insufficient} to provide enough extra squeeze of $O^{2-}$ ions to prevent lateral overswing of $3s$ electrons.

Next, add above the $\times$ mark in Fig. 13 one $O$ atom in the top terminal layer. It will 

\pagebreak

\includegraphics[width=5.05in]{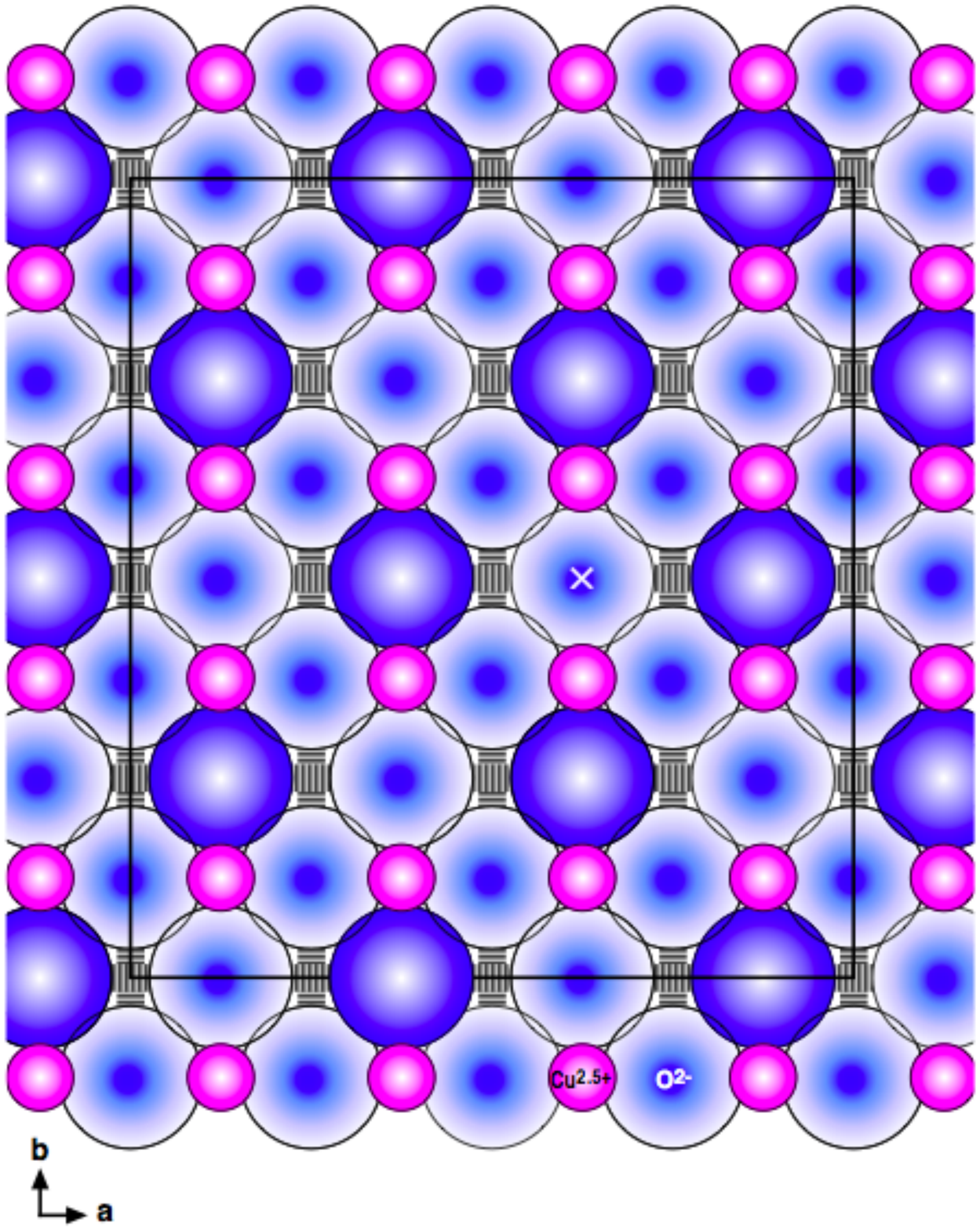}

\noindent FIG. 15. Upper $CuO_{2}$ plane in $YBa_{2}Cu_{3}O_{6.5}$: Each $O^{2-}$ ion with bold shading sits beneath an $O^{2-}$ ion in the oxygen chain above (along $+c$). All $O^{2-}$ ions are bracketed by a pair of copper ions $Cu^{2.5+}$, representing the time-average of a fluctuating $Cu^{2+}$-$Cu^{3+}$ pair. The $CuO_{2}$ plane is paved entirely by [$4Cu^{2.5+}$] copper rectangles. Their ionization and ensuing ion displacement creates enough extra squeeze of $O^{2-}$ ions to confine lateral oscillation of $3s$ electrons between $nnn$ $O^{2-}$ to wide electron tracks. This gives rise to superconductive connectivity, here with $T_{c} \simeq 57$ K at the onset of the first $T_{c}$-plateau. (The $O^{2-}$ with $\times$ mark prepares for changes in Fig. 16.)

\includegraphics[width=5.05in]{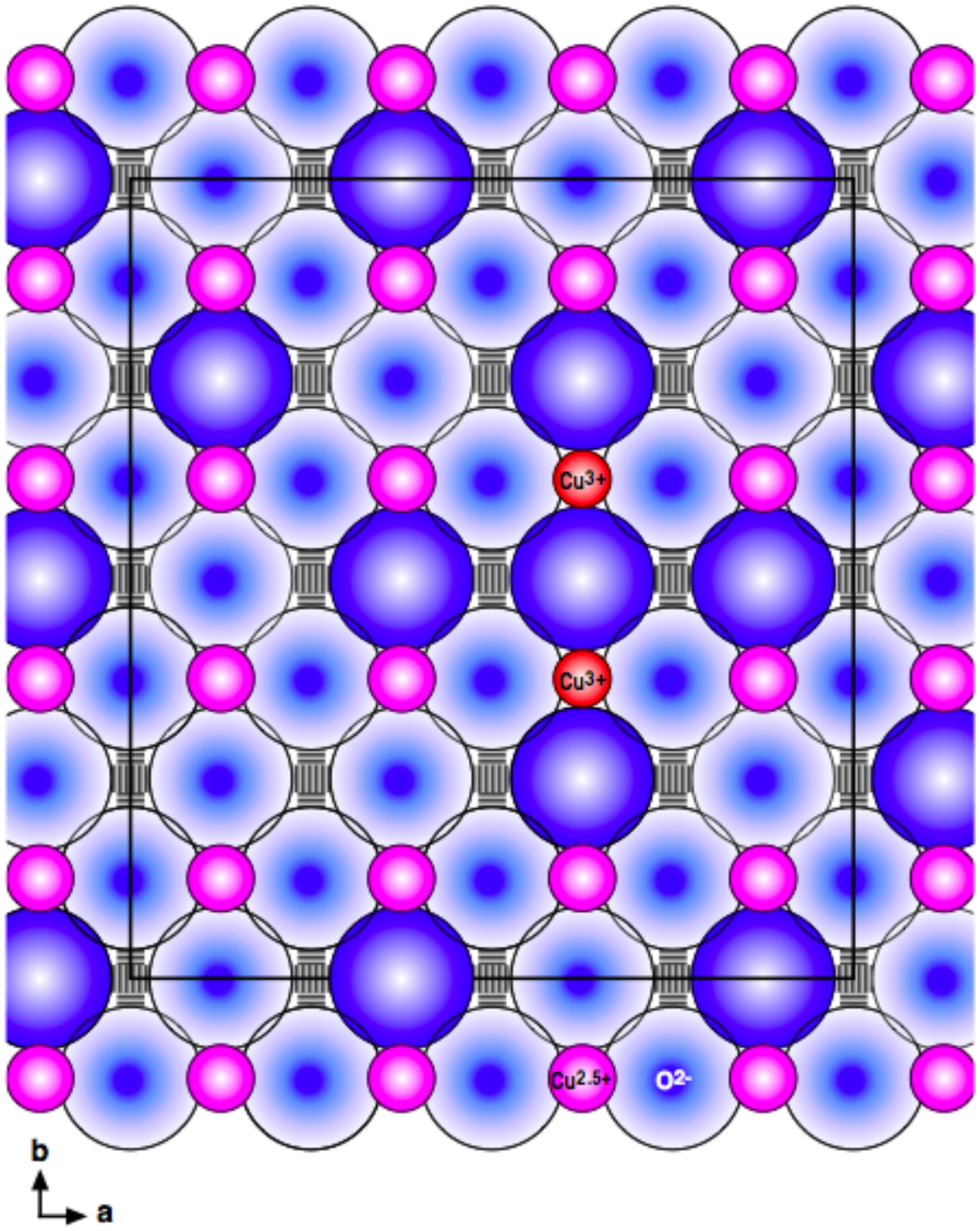}

\noindent FIG. 16. Upper $CuO_{2}$ plane in $YBa_{2}Cu_{3}O_{6+x}$ for $x = 1/2$ plus \emph{one} additional $O$ atom in the oxygen chain above the $\times$ mark in Fig. 15. Its ionization to $O^{2-}$ converts two $Cu^{2.5+}$ ions to $Cu^{3+}$. The ionization of the surrounding [$3Cu^{2.5+}$---$1Cu^{3+}$] and [$2Cu^{2.5+}$---$2Cu^{3+}$] copper rectangles is \emph{insufficient} to change the wide-track confinement of $3s$ electrons between $nnn$ $O^{2-}$ ions. This perpetuates the constant transition temperature of the first $T_{c}$-plateau.
\pagebreak

\noindent ionize to $O^{2-}$ by depriving two nearby $Cu^{2+}$ in the upper $CuO_{2}$ plane, marked by rings, of \emph{one} electron, creating a new $Cu^{2.5+}$ pair. The resulting situation is depicted in Fig. 14. Here we see that some $[4Cu^{2.5+}]$ rectangles have been formed. Their ionization causes enough ion displacement and extra squeeze of $O^{2-}$ ions to confine the lateral oscillation of $3s$ electrons to wide electron tracks. The result is a cross-shaped region of superconductivity with an arm-spread of $3a_{0}$ and a stem-length of $3b_{0}$. With increased oxygenation more such superconducting cross islands will emerge. It is a matter of percolation (not treated here) how these islands connect.

Enough thermal agitation---lattice vibrations in a crystal---destroy confinement of lateral $3s$ electron oscillation, leaving lateral overswing. It takes a certain amount of thermal energy, $\overline{\varepsilon} > {k_B}\overline{T}_{c}$, to destroy the lateral $3s$ confinement of a wide-track electron path, as in the [$4Cu^{2.5+}$], [$3Cu^{2.5+}$---$1Cu^{3+}$] and [$2Cu^{2.5+}$---$2Cu^{3+}$] copper rectangles. (Here the notation with overbar relates to a plateau and the hat below relates to a ramp.) This explains the \emph{constant} transition temperature of the first $T_{c}$-plateau,
\begin{equation}
{T_c}(x) = \overline{T}_{c} \, , \; 0.5 \, < \, x \, < \, 0.67,
\end{equation}
\noindent $\overline{T}_{c} \simeq 57$ K, where the $CuO_{2}$ plane is continguously paved by those wide-track copper rectangles. 

It takes a \emph{lesser} amount of thermal energy, $\hat{\varepsilon}(x) < \overline{\varepsilon}$, to destroy percolating connections of wide-track [$4Cu^{2.5+}$] copper rectangles (forming superconducting cross islands). The corresponding transition temperature $\hat{T}_{c}(x)$ is in proportion to the robustness of percolating connectivity of such copper rectangles, which in turn is proportional to their multitude (density in the $CuO_{2}$ plane). This qualitatively explains the linear rise of the transition temperature in the ramp range of oxygen enrichment, 
\begin{equation}
{T_c}(x) = \hat{T}_{c}(x) = 6(x - \frac{1}{3}) \, \overline{T}_{c} \, , \; 0.33 \, < \, x \, < \, 0.5.
\end{equation}

Figure 15 shows that for oxygen enrichment $x = 0.5$ the $CuO_{2}$ plane is paved entirely by [$4Cu^{2.5+}$] copper rectangles. Their ionization and the ensuing ion displacement creates enough extra squeeze of $O^{2-}$ ions to confine lateral oscillation of $3s$ electrons between $nnn$ $O^{2-}$ to wide electron tracks. This gives rise to superconductive connectivity, here with $T_{c} \simeq 57$ K at the onset of the first $T_{c}$-plateau.

Insertion of one $O$ atom in the oxygen chain above the $\times$ mark in Fig. 15 converts, after 

\pagebreak

\includegraphics[width=5.05in]{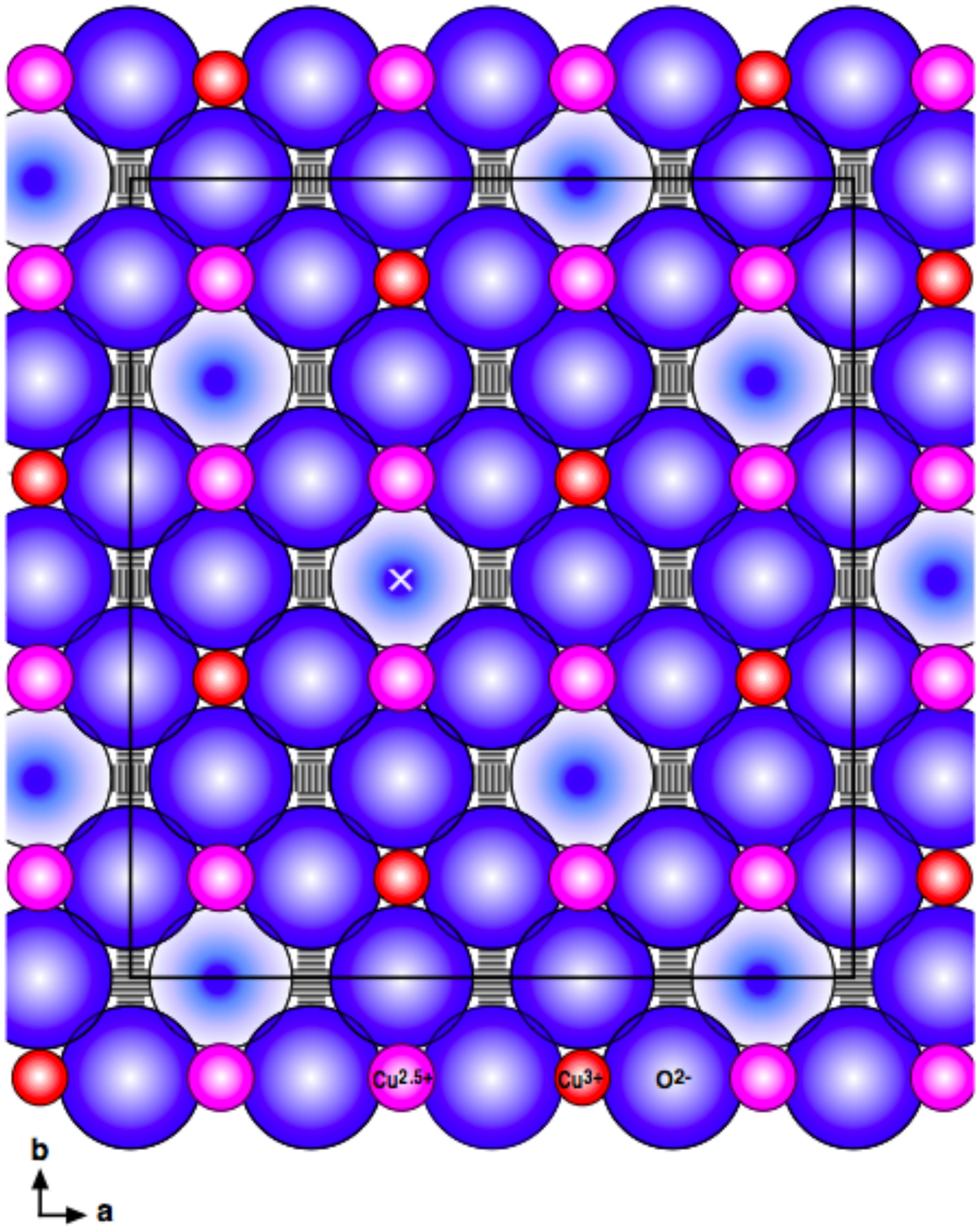}

\noindent FIG. 17. Upper $CuO_{2}$ plane in $YBa_{2}Cu_{3}O_{6.67}$: Each $O^{2-}$ ion with \emph{bland} shading sits beneath an \emph{empty} site in the oxygen chain above (along $+c$). Those $O^{2-}$ ions are bracketed by a pair of copper ions $Cu^{2.5+}$ of larger size than the $Cu^{3+}$ ions otherwise. The $CuO_{2}$ plane is paved by [$3Cu^{2.5+}$---$1Cu^{3+}$ and [$2Cu^{2.5+}$---$2Cu^{3+}$] copper rectangles. Their ionization and ensuing ion displacement is \emph{insufficient} to change the wide-track confinement of $3s$ electrons between $nnn$ $O^{2-}$ ions. (The $O^{2-}$ with $\times$ mark prepares for changes in Fig. 18.)

\includegraphics[width=5.05in]{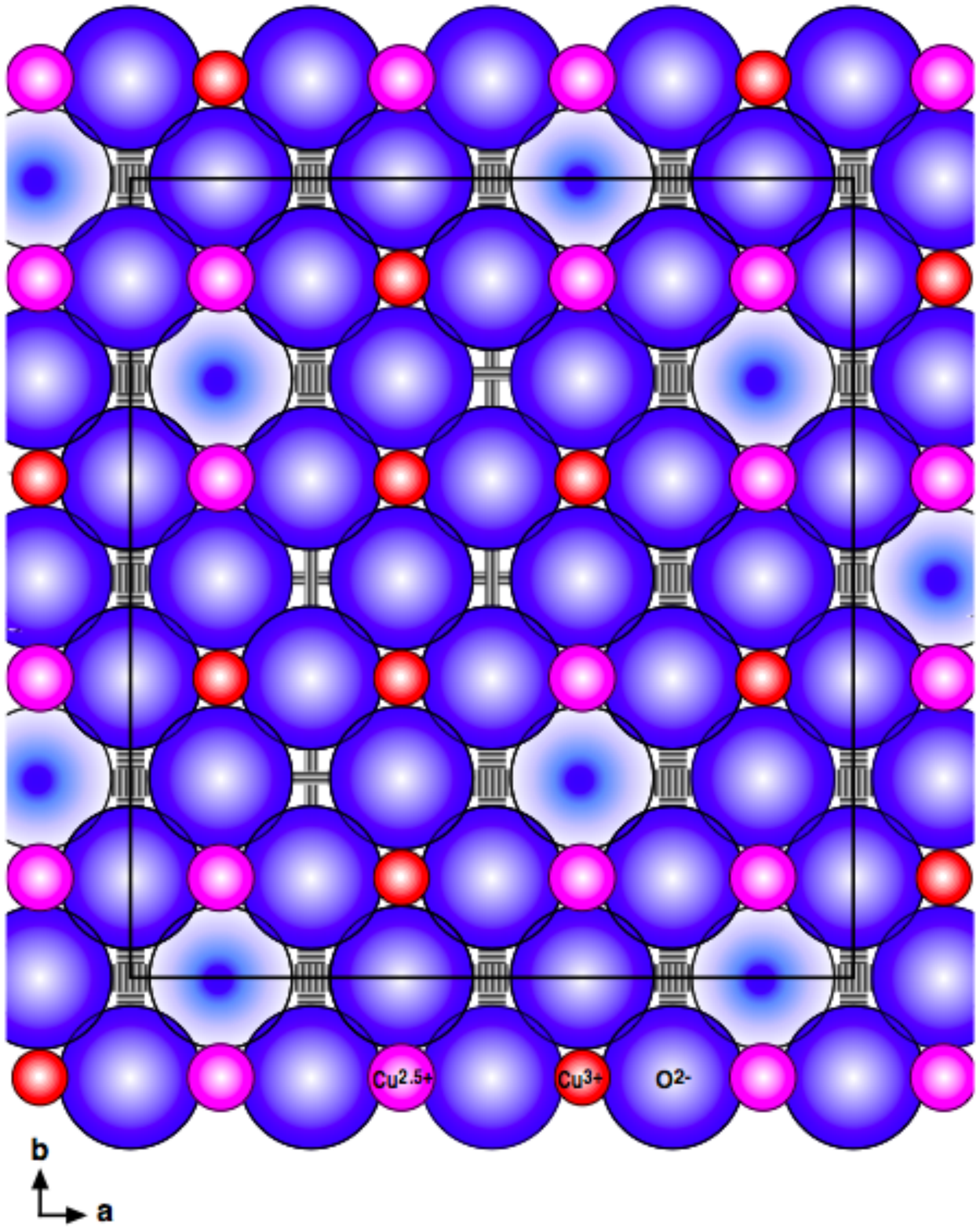}

\noindent FIG. 18. Upper $CuO_{2}$ plane in $YBa_{2}Cu_{3}O_{6+x}$ for $x = 2/3$ plus \emph{one} additional $O$ atom in the oxygen chain above the $\times$ mark in Fig. 17. Its ionization to $O^{2-}$ converts two $Cu^{2.5+}$ ions to $Cu^{3+}$. This creates four [$1Cu^{2.5+}$---$3Cu^{3+}$] copper rectangles with lateral $3s$ oscillation confined to narrow electrons tracks. They form a lightning-shaped superconducting path in a $1a_{0} \times 2b_{0}$ region. It signifies the onset of the second $T_{c}$-ramp.
\pagebreak

\noindent ionization to $O^{2-}$, the pair of bracketing $Cu^{2.5+}$ ions to $Cu^{3+}$ (see Fig. 16). The ionization of the surrounding [$3Cu^{2.5+}$---$1Cu^{3+}$] and [$2Cu^{2.5+}$---$2Cu^{3+}$] copper rectangles is \emph{insufficient} to change the wide-track confinement of $3s$ electrons between $nnn$ $O^{2-}$ ions. This perpetuates the constant transition temperature of the first $T_{c}$-plateau until the $CuO_{2}$ plane is completely paved, at $x = 2/3$, with a $\frac{2}{3}$-majority and, respectively, $\frac{1}{3}$-minority of those copper rectangles (see Fig. 17).

Adding one $O$ atom in the oxygen chain above the $\times$ mark in Fig. 17 converts, after ionization to $O^{2-}$, the bracketing $Cu^{2.5+}$ pair to $Cu^{3+}$. This creates four narrow-track [$1Cu^{2.5+}$---$3Cu^{3+}$] copper rectangles that form a lightning-shaped superconducting path across a $1a_{0} \times 2b_{0}$ region (see Fig. 18). It sets the stage for the second $T_{c}$-ramp. Further oxygen enrichment, $0.67 < x < 0.83$, creates more narrow-track [$1Cu^{2.5+}$---$3Cu^{3+}$] rectangles in the $CuO_{2}$ plane. The increasing robustness of their percolating connectivity is reflected by the steady rise of superconductive transition temperature (second ramp). This continues until $x = 5/6 \simeq 0.83$ when the $CuO_{2}$ plane is contiguously paved by a $\frac{2}{3}$-majority of such [$1Cu^{2.5+}$---$3Cu^{3+}$] rectangles, except for a $\frac{1}{3}$-minority of isolated wide-track [$2Cu^{2.5+}$---$2Cu^{3+}$] pockets, shown in Fig. 19. The narrow-track [$1Cu^{2.5+}$---$3Cu^{3+}$] copper rectangles form swaths of superconducting pathways through the $CuO_{2}$ plane. (If you run off the right margin of Fig. 19 when tracing the narrow-track connectivity, then continue at the left margin which extends periodically.)

Lastly, add one $O$ atom in the oxygen chain above the $\times$ mark in Fig. 19. It converts, after ionization to $O^{2-}$, the bracketing $Cu^{2.5+}$ pair to $Cu^{3+}$, and thereby converts two flanking wide-track [$2Cu^{2.5+}$---$2Cu^{3+}$] copper rectangles to two additional narrow-track rectangles of the [$4Cu^{3+}$] type.  This perpetuates the constant transition temperature $T_{c} \simeq 90$ K of the second $T_{c}$-plateau, $0.83 < x < 1$, until the $CuO_{2}$ plane is completely paved with [$4Cu^{3+}$] rectangles in $YBa_{2}Cu_{3}O_{7}$.

Two final remarks seem in order: (1) It is a direct consequence of the crystal structure of $YBa_{2}Cu_{3}O_{6+x}$ that both the tetragonal $\to$ orthorhombic phase transition (at $x = 1/3$) and the onset of $T_{c}$ ramps and plateaus (at $x_{n} = n/6, \; n=2,...,5$) occur at simple ratios of oxygen enrichment. (2) The confinement of lateral $3s$ electron oscillation between $nnn$ $O^{2-}$ ions to either wide or narrow electron tracks seems to be quantized and calls for confirmation with quantum-mechanical methods.

\pagebreak

\includegraphics[width=5.05in]{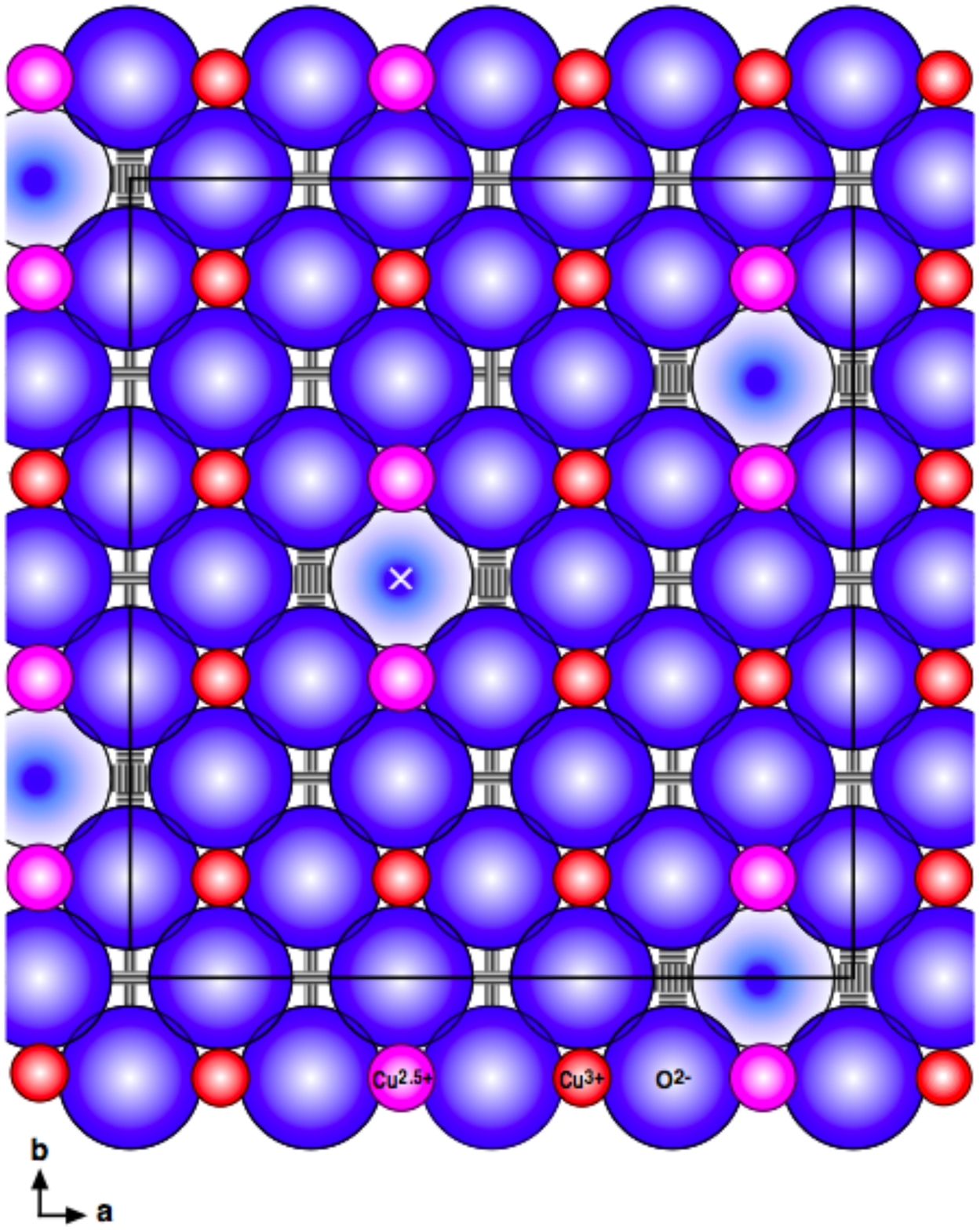}

\noindent FIG. 19. Upper $CuO_{2}$ plane in $YBa_{2}Cu_{3}O_{6.83}$: Each $O^{2-}$ ion with bland shading sits beneath an \emph{empty} site in the oxygen chain above (along $+c$). Those $O^{2-}$ ions are bracketed by a pair of $Cu^{2.5+}$ copper ions of larger size than the $Cu^{3+}$ ions otherwise. The $CuO_{2}$ plane is paved by [$2Cu^{2.5+}$---$2Cu^{3+}$] and [$1Cu^{2.5+}$---$3Cu^{3+}$] copper rectangles. The ionization and ensuing ion displacement of the latter causes narrow-track confinement of $3s$ electrons between $nnn$ $O^{2-}$ ions. The [$1Cu^{2.5+}$---$3Cu^{3+}$] copper rectangles form a percolating path throughout the $CuO_{2}$ plane. (To extend the right margin of the figure, when tracing narrow-track connections, continue at the left margin which extends periodically.) The $O^{2-}$ with $\times$ mark prepares for changes in Fig. 20.
\pagebreak

\includegraphics[width=5.05in]{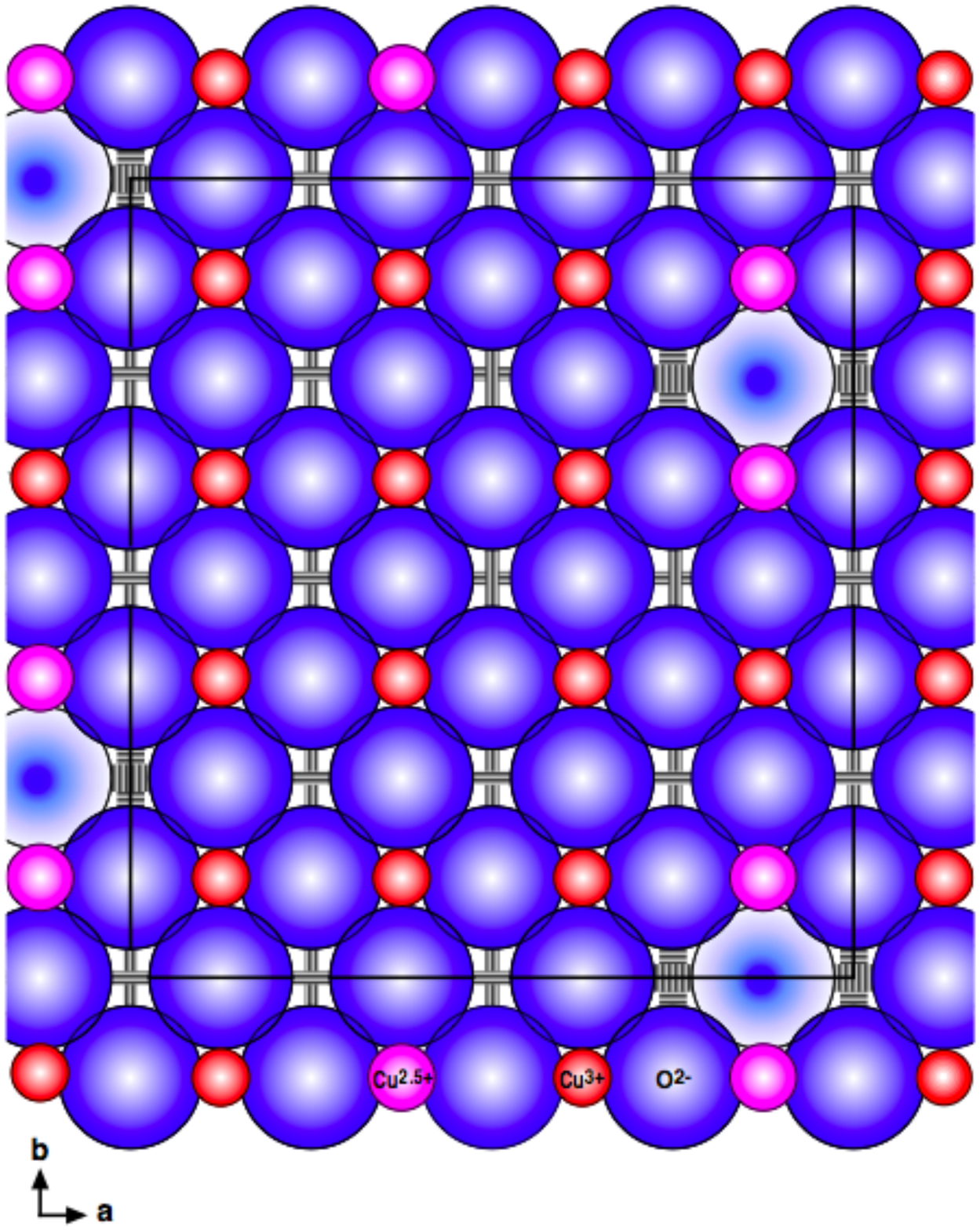}

\noindent FIG. 20. Upper $CuO_{2}$ plane in $YBa_{2}Cu_{3}O_{6+x}$ for $x = 5/6$ plus \emph{one} additional $O$ atom in the oxygen chain above the $\times$ mark in Fig. 19. Its ionization to $O^{2-}$ converts two $Cu^{2.5+}$ ions to $Cu^{3+}$. This creates some [$4Cu^{3+}$] copper rectangles in the neighborhood.  They have the same narrow-track confinement of $3s$ electrons between $nnn$ $O^{2-}$ ions as the contiguous [$1Cu^{2.5+}$---$3Cu^{3+}$] copper rectangles. This perpetuates the constant transition temperature of the second $T_{c}$-plateau.
\pagebreak

\centerline{ \textbf{ACKNOWLEDGMENTS}}

\noindent I thank Duane Siemens for stimulating discussions and Preston Jones for help with LaTeX.

\end{document}